\documentclass[preprint,12pt]{elsarticle}

\usepackage{amssymb}
\usepackage{lscape}
\usepackage[T1]{fontenc}
\usepackage{amsmath}

\newcounter{bla}

\journal{Computer Physics Communications}

\begin{document}

\begin{frontmatter}



\title{Automatic classification of nuclear physics data via a Constrained Evolutionary Clustering approach}


\author[a]{D. Dell'Aquila\corref{author}}
\author[b,c]{M. Russo}

\cortext[author] {Corresponding author.\\\textit{E-mail address:} daniele.dellaquila@irb.hr}
\address[a]{Ru{\dj}er Bo\v{s}kovi\'c Institute, Department of Experimental Physics, Bijeni\v{c}ka 54, Zagreb, HR-10000, Croatia}
\address[b]{Dipartimento di Fisica e Astronomia, Universit\`a degli Studi di Catania, 95235 Catania, Italy}
\address[c]{INFN-Sezione di Catania, 95235 Catania, Italy}

\begin{abstract}
This paper presents an automatic method for data classification in nuclear physics experiments based on evolutionary computing and vector quantization. The major novelties of our approach are the fully automatic mechanism and the use of analytical models to provide physics constraints, yielding to a fast and physically reliable classification with nearly-zero human supervision. Our method is successfully validated by using experimental data produced by stacks of semiconducting detectors. The resulting classification is highly satisfactory for all explored cases and is particularly robust to noise. The algorithm is suitable to be integrated in the online and offline analysis programs of existing large complexity detection arrays for the study of nucleus-nucleus collisions at low and intermediate energies.
\end{abstract}


\end{frontmatter}

\section{Introduction}
\label{sec:sec1}
Nuclear physics experiments significantly rely on data classification, i.e. the grouping of data into meaningful physics classes, to reconstruct nucleus-nucleus collision events and enable the exploration of the underlying physics. In studies that exploit the detection of charged particles, the classification problem is often that of identifying charge ($Z$) and mass ($A$) of detected ions. This process is usually indicated as \emph{particle identification}. To this end, a number of detection systems capable to record information useful to the classification process have been developed in the last decades \cite{Pouthas95,Pagano12,Wuenschel09,Davin01,Wallace07,DellAquila19,Bougault14,Acosta16,Lukasik13,MarquinezDuran14}. A quite common strategy consists in the use of detector arrays based on stacks of detection layers through which the particle penetrates before being completely stopped. In similar arrays, if organized in a 2D correlation plot, data recorded by pairs of independent layers assemble into bi-dimensional non-overlapping clusters, each representing a certain ($Z$,$A$) class. To this extent, the problem of nuclear physics data classification is equivalent to the extraction of clusters in a bi-dimensional space.

Numerous algorithms for Cluster Analysis (CA) or Vector Quantization (VQ) have been proposed in the literature so far, achieving a noticeable success in standard partitioning problems and being focused on obtaining clusters of nearly equal dispersion (see for example \cite{Patane01,Patane02,Baraldi02,Frigui99}). However, the clusterization process in nuclear physics data is made more difficult by the large variability of size and dispersion of the various clusters \cite{Benkirane95}. Because of these unique features, an optimal solution according to the CA/VQ approach, where a good equalization of the content of each cluster is obtained, might not be directly applicable to nuclear physics classification problems, where an acceptable physical solution usually contains clusters with strongly unbalanced distortions. For this reason, only a reduced number of works have been previously carried out attempting to use CA/VQ methods in nuclear data classification problems. For example, fuzzy c-means algorithms have been successfully applied to the identification of particles in nucleus-nucleus collisions, but their use was restricted, exclusively, to cases characterized by the presence of few clusters with nearly equal distortion \cite{Wirth13}.

Several studies have been instead focused on the classification of nuclear physics data in more general cases. Among those, image processing techniques are widely applied. For example, unsupervised learning approaches exploiting contextual image segmentation methods \cite{Benkirane95}, neural networks \cite{Alderighi01} or spatial density analysis \cite{Alderighi01b} led to quite satisfactory classifications. However, these methods were conceived to classify exclusively $Z$-values, thus being not particularly suitable for the vast majority of modern high-resolution experiments, where $A$-classification is often a crucial requirement \cite{Bougault14}. In more recent times, another automatic classification method was proposed in Ref.~\cite{Morach08}. This method allowed a good extraction of clusters, but the procedure does not comprise an explicit link between extracted clusters and physically meaningful classes, thus requiring a significant human supervision, especially in the analysis of data produced by large detection arrays.

Because of the above discussed limitations, the vast majority of the approaches commonly used for the classification of data in nuclear physics experiments involve human-supervised techniques. In similar approaches, the operator manually extracts information by visually inspecting bi-dimensional distributions of data, which is then used as input for supervised learning procedures. With this respect, artificial neural networks have been proposed \cite{IaconoManno00}. More usually, error minimization procedures based on mathematical models \cite{TassanGot02,LeNeindre02}, which contain $Z$ and $A$-values explicitly, are instead preferred. The latter allow to manually extract information only for a reduced number of clusters, while the resulting classification can be meaningfully extended to any possible ($Z$,$A$) ion by model extrapolation. An additional reduction in the required information can be finally obtained if one follows the procedure suggested in Ref.~\cite{Gruyer17}.

However, despite the significant effort poised to minimize human supervision, obtaining a physically meaningful data classification in nuclear physics experiments is still a quite repetitive and time consuming task for the operator, especially in the case of modern detection arrays, where the number of individual bi-dimensional plots to inspect ranges from few hundreds to thousands. As an example, depending on the number of bi-dimensional assembly to classify, the time required to an operator to perform a similar task ranges from days to months. In this framework, it is clear that new fully-automatic methods for data classification are highly required.

In this paper, we present an innovative approach to nuclear physics data classification that allows to reduce the task to few minutes or hours of machine time with nearly-zero supervision by the operator. Our approach involves Evolutionary Computing (EC) and CA/VQ and is based on an two-level search for the optimal solution of the data clusterization problem. The upper level consists in a global search operated by an EC algorithm that treats each solution as an individual of a given population and applies some suitable evolutionary criteria. In our EC approach, an individual is encoded according to a given choice of functional parameters, which constrain a physically meaningful model for the description of bi-dimensional clusters\footnote{For the present work, the model proposed in Refs.~\cite{TassanGot02,LeNeindre02} has been used. However, our algorithm is fully general and can be applied exploiting any given model with an arbitrary number of parameters to adjust.}, and a number of ($Z$,$A$) physics classes. The lower level, used as a local \emph{hill-climbing} operator for the EC process, performs a fast local search through a suitable VQ algorithm that exploits the resulting EC individual as the initial codebook for the optimization problem. The major novelty with respect to previously published CA/VQ methods is that the codebook has a physical meaning and resulting solutions are immediately applicable to the data classification with nearly-zero effort required to the operator. In addition, our algorithm is fully automatic as no a priori information is required to the experimenter.

Since the proposed methodology is highly interdisciplinary, in order to effectively drive the reader through the paper, we provide with an introduction of all individual research fields that cooperate in our algorithm. The paper is organized as follows: Section~\ref{sec:sec2} gives a more quantitative description of the classification problem studied in this paper, Section~\ref{sec:sec3} is concerned with a description of the programming techniques used in our study, i.e. EC and CA/VQ, Section~\ref{sec:sec4} provides a detailed description of the algorithm, in Section~\ref{sec:sec5} we test the performance of the algorithm in a typical experimental case, in Section~\ref{sec:sec6} we compare our methodology with previously published approaches and, finally, Section~\ref{sec:sec7} reports conclusions and possible future perspectives of our work.

\section{Experimental context}
\label{sec:sec2}
Charged particles are among the most frequent products of a nucleus-nucleus collision. At low and intermediate incident energies ($\leq200$ MeV/A) they consist almost exclusively of heavy ions ($Z\geq1$). Their number in a typical collision event varies depending on the incident energy and the size of the nuclei involved in the collision and can range from a few units to more than $50$. To meet the requirements of modern nuclear physics studies, state-of-the-art apparatus for the detection of charged particles have reached an extremely high level of accuracy in identifying a large variety of ions.

\begin{figure}[t]
	\centering
	\includegraphics[scale=0.5]{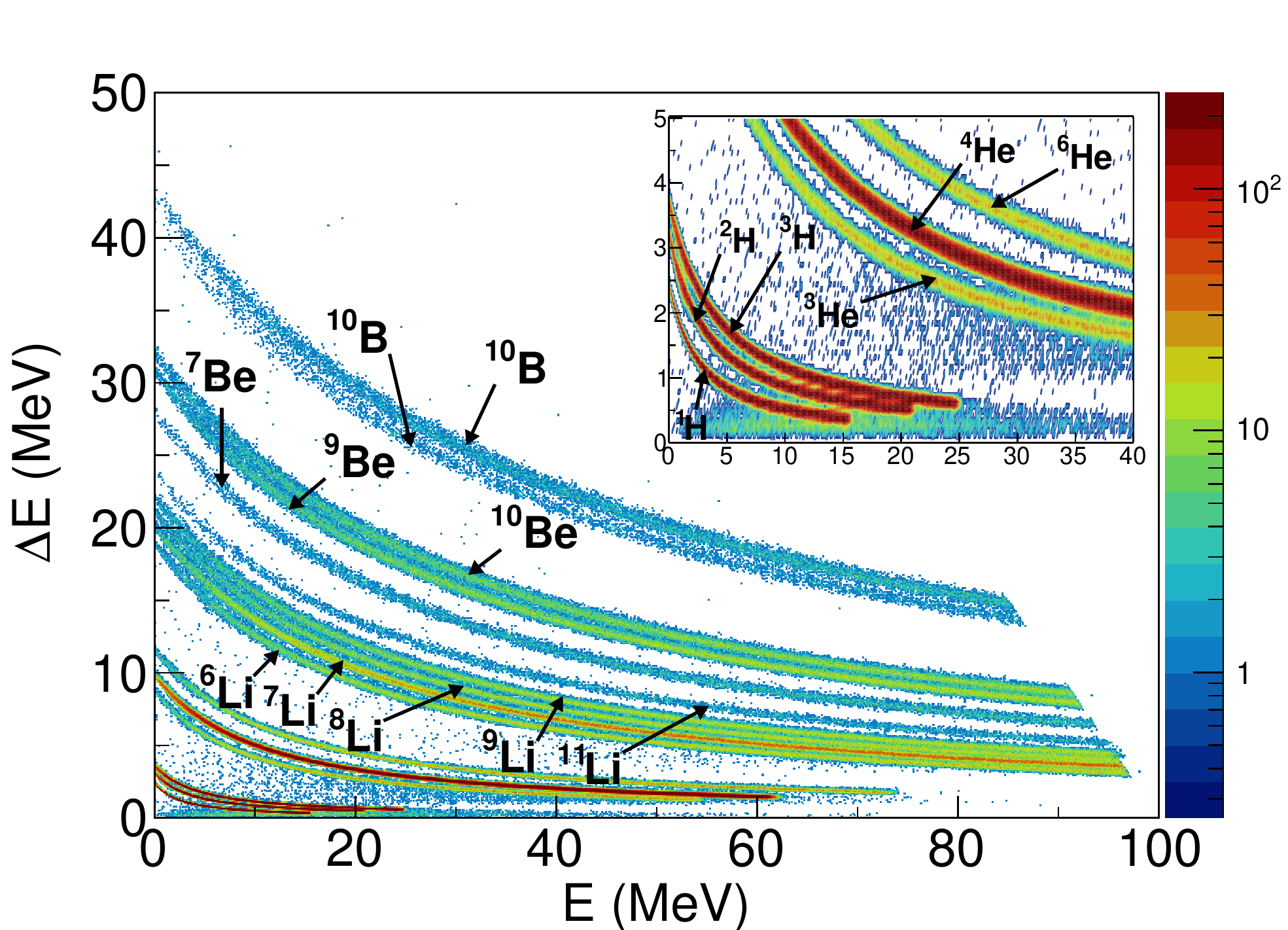}
	\caption{A typical plot obtained by correlating the energy signals of two longitudinally stacked detectors used as $\Delta$E-E telescope. Data is obtained by means of the GEANT4 toolkit \cite{Agostinelli03} for a Silicon ($65$ $\mu$m)-Silicon telescope. In such a bi-dimensional representation, data group into clusters, each corresponding to a given ($Z$, $A$) ion. Colors define the statistics of counts as indicated by the color scale. The insert represents a zoom of the $1\leq Z\leq 2$ region. Labels indicate loci corresponding to the various ions.}
	\label{fig:fig01}
\end{figure}
Two numerical quantities are typically used to identify an ion: its number of protons, often called \emph{charge} and indicated with $Z$, and its total number of nucleons, i.e. the sum of the number of its protons and neutrons, called \emph{mass} and indicated with $A$. Because the energy deposition of a heavy ion in a given material is a well-known function of its energy (i.e. its velocity), the properties of the material and the nature of the ion (i.e. $Z$ and $A$-values) \cite{Livingston37}, one can identify an ion produced in a collision event if its velocity and its energy loss in a layer of a material of given properties are known. This is the principle of the particle identification technique called $\Delta$E-E, which is probably the most widely used approach to identify charged particles in a nucleus-nucleus collision at intermediate and low energies. The practical implementation of the $\Delta$E-E technique relies on the use of so-called telescopes, i.e. stacks of longitudinally arranged detectors, each with an independent readout. Two detection stages are typically sufficient to this end, but configurations with more than two stages have also been proposed to account for the need for a particularly large dynamic range in a single experiment (see e.g. \cite{Bougault14,Acosta16,Wallace07}). Let us assume, for the sake of clarity, that a telescope composed of two detection stages is used and that the particle has sufficient energy to pass through the first stage, being stopped in the second stage\footnote{The approach is more general and can be applied to any array of longitudinally stacked detectors if signals produced by pairs of independent layers are considered; to this end, the hypothesis that the particle is fully stopped in the second of those layers is required.}. In a similar telescope, the signal recorded by the first detection stage, associated to the energy deposited by the particle in the detector volume, corresponds only to a portion of the total kinetic energy $E_{kin}$ of the particle and can be therefore indicated as $\Delta E$. Since the particle is stopped in the second detection stage, the signal produced by the latter will complement the kinetic energy of the particle, i.e. $E_{kin}=\Delta{E}+{E}$. If the first stage is sufficiently thin, the following equation can be easily derived from the non-relativistic formalism introduced in \cite{Livingston37}:
\begin{equation}
  \label{eq:eq1}
  -\frac{dE}{dX}=\frac{4\pi e^4 Z^2}{m_ev^2}NB \approx \frac{\Delta E}{\Delta X} \Rightarrow {\Delta E} \propto Z^2 A \frac{1}{E}
\end{equation}
where $e$ is the elementary electric charge, $m_e$ is the mass of the electron, $v$ is the velocity of the incident particle, $N$ is the number of atoms per $cm^3$ in the material, $I$ is the average excitation potential of the atoms in the material and $B$ is a dimensionless quantity weakly dependent on the velocity of the particle and on the properties of the material. In the last term, we replaced $E_{kin}$ with $E$, having $E_{kin}=\Delta{E}+E\approx E$ and we considered the approximation $B\approx const$. It is important to point out that these approximations and the simplified formalism derived from \cite{Livingston37} are adequate for the purpose of this section, a fully accurate formalism will be instead used to derive the classification algorithm of Section~\ref{sec:sec4}. From the last term of eq.~\ref{eq:eq1}, one can state that pairs of charged particle signals recorded by a telescope occupy hyperbolic-like loci in the ($E$,$\Delta E$) plane depending on their ($Z$,$A$), values. Loci are not equally spaced as the dependency is quadratic on $Z$ and linear on $A$. This can be observed in the plot shown on Fig.~\ref{fig:fig01}, where data simulated by using the GEANT4 toolkit \cite{Agostinelli03} for a typical Silicon-Silicon telescope are shown. To produce the data in figure, we have considered, for simplicity, a uniform energy distribution in the range $[0,100]$ MeV for all emitted particles, discarding the signals of particles not fully stopped in the second detection stage. The $Z$-value range $1\leq Z \leq 5$ was taken into consideration, and a realistic $A$ distribution for each $Z$-specie was introduced. By visually inspecting the bi-dimensional distribution, one can immediately observe that $5$ different areas are populated, corresponding to the $5$ $Z$-values considered in the example and representing H, He, Li, Be, B. As it can be more accurately seen in the insert, which shows a zoom of loci corresponding to $Z=1$ and $Z=2$, $Z$-lines are additionally separated into $A$-clusters, each corresponding to a different isotope of the particular $Z$-specie. Because the latter are intrinsically less separated than $Z$-lines, as a result of the linear dependence on $A$ given by eq.~\ref{eq:eq1}, their identification is usually more challenging. In a typical experiment, different $Z$ and $A$ distributions could be recorded by different detectors within the same setup, depending on their geometrical position with respect to the accelerated beam. Another interesting feature observed in Fig.~\ref{fig:fig01} is the difference in population of the various clusters. They reflect the statistics of production of different ions emitted in the nuclear collision. In the example of Fig.~\ref{fig:fig01}, where colors indicate the content of each individual bin, we have chosen a distribution similar to that observed in $^{9}$Li+$^{6}$Li, $^{9}$Li+$^{7}$Li and $^{9}$Li+$^{19}$F collisions at $65$ MeV incident energy. It is evident, for instance, how clusters corresponding to $Z=1$ are particularly more populated that $Z=5$ ones. The dispersion of each cluster around its mean value differs also very significantly cluster-by-cluster. For example, lines associated to $Z=1$ isotopes, i.e. $^{1}$H, $^{2}$H and $^{3}$H, well-visible in the insert, have an average dispersion of about $100$ keV FWHM, while that associated to $^{10}$B results in an average dispersion of the order of $600$ keV. These values, that usually increase for increasing $Z$-values, are characteristic of the process of interaction of the radiation with the matter and are therefore affected by the effective thickness of the $\Delta E$ detector as well as by the uniformity of the detector itself. In addition, the thickness of the $\Delta E$ detector defines the absolute position of clusters in the ($E$,$\Delta E$) plane, which is typically different detector-by-detector.

To summarize, the problem of charged particle identification with longitudinally stacked detectors consists in recognizing the ($E$,$\Delta E$) pairs that belong to the same physical class and to link them to a given ($Z$,$A$) ion. Such a task is therefore equivalent to a classification problem. The most relevant features of the bi-dimensional clusters to extract are: (i) number of clusters and their $Z$ and $A$ values are different for each experiment and often detector-by-detector, (ii) statistics within each cluster are significantly different to each other, (iii) clusters have non-equal dispersions, (iv) no reasonable hypothesis can be made regarding the absolute position of clusters in the ($E$,$\Delta E$) plane.

\section{Description of the programming techniques}
\label{sec:sec3}
The newly proposed methodology comprises two different programming techniques that cooperate to obtain, in a fully automatic way, a solution to the proposed classification problem: EC and CA/VQ. In this section, we provide an introduction on the most salient concepts of EC and CA/VQ.
\subsection{Evolutionary computing}
\label{sec:sec3.1}
Evolutionary computing is a scientific field that concerns with the resolution of optimization problems through the application of concepts derived from the natural world \cite{Koza92}. Nowadays, EC is applied to numerous domains of science \cite{Gotmare17,Darwish19,Russo16}. A very frequent scheme, which is derived from the Darwinian evolutionary theory, can be schematically described in the following way:
\begin{enumerate}
	\item A set of possible solutions to the optimization problem, encoded according to a predefined scheme, is generated (often randomly). Each of such solutions is called \emph{individual}, while a set of individuals forms a \emph{population}.
	\item A numerical value, called \emph{fitness}, is associated to each individual. The fitness quantifies how much a given solution is \emph{optimal} to the problem to solve. The higher is the fitness associated to an individual the more promising is the individual itself. This is a crucial quantity for the success of the optimization procedure.
	\item Until a predefined convergence criterion is reached, the following steps are iterated:
	\begin{enumerate}
		\item Some individuals are selected (\emph{parents}) to be used as a starting point for the generation of new individuals (\emph{offsprings}).
		\item Offsprings are obtained through a suitable mechanism of parents encoding recombination (\emph{crossover}). In this phase, the chromosomes of the parents, i.e. their encoding, are suitably combined to generate new individuals. A valid crossover should produce individuals whose genetic code is, to some extent, similar to that of the parents. Crossover is usually followed by a random variation, with low probability, of some portions of the derived encoding. Such a process is called \emph{mutation} and has a crucial importance as it allows to introduce missing genetic code and to keep genetic diversity in the population. The fitness is finally calculated for all newly obtained individuals.
		\item  Some offsprings live sufficiently long to replace other pre-existing individuals.
	\end{enumerate}
\end{enumerate}

The mechanism of fitness improvement typical of EC is a result of the implementation of suitable selection criteria. This is mandatory in order to enhance the performance of the algorithm with respect to purely Monte Carlo codes. As an example, the initial choice of the parents of step (a) can be affected by the fitness of the available individuals, i.e. parents can be chosen according to a probability distribution that favors the extraction of high-fitness individuals. In a similar way, the \emph{replacement} criterion used to introduce newely generated offsprings in the population can account for the fitness of pre-existing individuals. The latter are usually selected among those rejected in step (a). To this end, deterministic tournaments between pairs of individuals or fitness-based stochastic criteria are often used. The replacement of individuals is generally required in order to maintain constant the total number of individuals in the population.

EC algorithms are strongly CPU-oriented; this makes it crucial to allocate most of CPU resources towards more promising individuals, even if a certain CPU quota has to be ensured to all individuals in the population. A similar allocation of available resources is intrinsic in the selection process operated by EC.

Another fundamental aspect is the so-called \emph{premature convergence}. It consists in the convergence of the algorithm towards a local maximum of the fitness function and often negatively affects the capabilities of EC of obtaining satisfactory results. As shown in previous studies, see e.g. Ref.~\cite{Russo16}, the problem of premature convergence can be contrasted by subdividing the population into multiple sub-populations. In a similar way, even if a sub-population has reached a premature convergence, with a consequent loss of genetic diversity, it is extremely unlikely that all sub-populations simultaneously converge towards the same individual. In addition, \emph{migration} plays a fundamental role in this context. If a sub-population has converged towards a local maximum, thus stopping to improve individuals, injecting an individual from another independent sub-population might result in restarting the evolution for the prematurely converged sub-population.

The global search of EC is generally extremely powerful to obtain good solutions that are close to a maximum of the fitness function but is rather slow for the search for the maximum itself. Local search techniques are instead suitable for the fast determination of the closest local maximum. For this reason, one or more hill-climbing operators, devoted to a fast local search in the proximity of a maximum, are sometimes introduced in EC to significantly speed-up the determination of the maximum for the fitness function.

\subsection{Clustering analysis and vector quantization}
\label{sec:sec3.2}
Clustering is an important instrument in many scientific disciplines. The partitioning approach known as VQ \cite{Gersho92} consists in the derivation of a set (\emph{codebook}) of reference or prototype vectors (\emph{codewords}) from a given data set. In a similar way, each subset of vectors (\emph{patterns}) belonging to the original dataset is represented uniquely by one codeword. Clusters can be easily extracted based on their proximity to the available codewords. While VQ is concerned with findings a codebook to represent the original multi-dimensional dataset as well as possible, CA is conceived as the problem of identifying clusters of data, regardless the determination of a codebook. Codewords are determined by a procedure that consists in the minimization of an objective function (\emph{distortion}) representing the \emph{quantization error} (QE) \cite{Hofmann97}. A widespread accepted classification scheme distinguishes between $K$-means and competitive learning. Clustering algorithms belong to the first of those classes \cite{Linde80,Patane01} and are based on the minimization of the average distortion through a suitable choice of codewords. In approaches belonging to the second category, a codebook is instead obtained as a consequence of a competition process between codewords \cite{Kohonen89}. 

Quantitatively, the objective of standard VQ consists in the representation of a given set of vectors ${\bf x}\in X\subseteq \Re^k$ through a set, $Y=\{{\bf y}_1,...,{\bf y}_{N_C}\}$, of $N_C$ reference vectors in $\Re^k$. In this definition, the vectors ${\bf y}_i$ represent the codewords and $Y$ is the codebook. A VQ can be therefore represented by a function $q: X \longrightarrow Y$. The determination of $q$ allows to obtain a partition ${\cal S}$ of the original dataset $X$ constituted by $N_C$ subsets, $S_i$, called \emph{cells}:
\begin{equation}
  {\cal S}=\{S_i;\:\:i=1,\ldots,N_C\}
  \label{eq:partition}
\end{equation}
Where each cell $S_i$ is defined by the following equation:
\begin{equation}
 {S_i=\{{\bf x}\in X: q({\bf x})={\bf y}_i\}}
 \label{eq:insiemi}
\end{equation}

\subsubsection{The Quantization Error}
QE is the value deduced by $d({\bf x},q({\bf x}))$, being $d$ a generic distance operator between vectors defined in $X\times Y$. Several functions are conventionally used for distortion measurement \cite{Linde80}. For the purpose of this work, as it will be discussed in detail in Section~\ref{sec:sec4}, the scheme introduced by eq.~\ref{eq:insiemi} has been modified to be suitable for our classification problem, where codewords are represented by hyperbolic-like curves of the type of eq.~\ref{eq:eq1} and the distance operator $d$ is defined accordingly. However, the following discussion is valid without any lack of generality.

The performance of a given quantizer $q$ is usually evaluated through the Mean QE (MQE). When $X$ is constituted by a finite number ($N_P$) of patterns, MQE is generally given by:
\begin{equation}
  {\rm MQE} \equiv D(Y,{\cal S})= \frac{1}{N_P}\sum_{i=1}^{N_C}D_i
  \label{eq:dist_cod}
\end{equation}
where $D_i$ is the total distortion of the $i$-th cell, being it defined by the following equation:
\begin{equation}
  D_i = \sum_{n:{\bf x}_n \in {\cal S}_i} d({\bf x}_n,q({\bf x}_n))
  \label{eq:tot_cella}
\end{equation}
Equation \ref{eq:dist_cod} shows that MQE is equivalent to a function ($D$) of the codebook $Y$ and the corresponding partition $\cal S$.

The core of a CA/VQ algorithm consists in the application of two important conditions that are used for the calculation of the optimal partition (when the codebook is fixed) and the optimal codebook (when the partition is fixed) \cite{Linde80}:
\begin{itemize}
	\item \emph{Nearest Neighbor Condition (NNC)}. The NNC consists in assigning the nearest codeword, according to the meaning given by the metric $d$, to each pattern in the original dataset $X$. For a given codebook $Y$, the following partition is thus identified by the NNC:
	\begin{equation}
	\bar{S}_i=\{{\bf x}\in X : d({\bf x},y_i)\leq d({\bf x},y_j) \ \ \forall y_j\in Y)\}
	\label{eq:voronoi}
	\end{equation}
	The set ${\bar{S}_i}$ defined by the NNC corresponds to the so-called \emph{Voronoi Partition} \cite{Gersho92} of the original dataset and is usually indicated with the symbol ${\cal P}(Y)=\{\bar{S}_1, \cdots, \bar{S}_{N_C}\}$. ${\cal P}(Y)$ corresponds to the optimal $X$ partition, given the codebook $Y$ \cite{Linde80}.
	\item \emph{Centroid Condition (CC)}. The CC is concerned with the procedure of finding the optimal codebook for a certain partition ${\cal S}$ of the original dataset $X$, i.e. to determine the centroid $\bar{x}$ of each individual cell $S_i$, according to the given metric. The corresponding codebook will be then defined by grouping all centroids $\bar{x}(S_i)$:
	\begin{equation}
	  \bar{X}({\cal S})\equiv\{\bar{x}(S_i);\:\:i=1,...,N_C\}
	  \label{eq:centroids}
	\end{equation}
\end{itemize}

\subsubsection{LBG and $K$-means}
LBG is an iterative algorithm that, $N_C$ being fixed, for each iteration produces a quantizer $q$ better than or equal to the one obtained in the previous iteration. This approach is practically equivalent to that of the traditional $K$-means \cite{Jain00}. The steps through which LBG develops can be schematically described as follows:
\begin{enumerate}
	\item \emph{Initialization}: in this phase, an initial codebook is chosen according to a given approach, often randomly (see e.g. \cite{Linde80}).
	\item \label{lbg:partition_calculation}\emph{Partition calculation}: Given the codebook determined in the previous step, the related Voronoi Partition, (eq. \ref{eq:voronoi}) is calculated according to the NNC.
	\item \emph{Termination condition}: The MQE at the current iteration $D_{\rm curr}$ is compared with the one obtained in the previous iteration $D_{\rm prev}$. If the ratio	$|D_{\rm prev}-D_{\rm curr}|/D_{\rm prev}$ is less than a prefixed threshold ($\varepsilon$) then the algorithm ends; otherwise, it continues with the next step;
	\item \emph{Codebook calculation}: By using the partition calculated in step~\ref{lbg:partition_calculation}, a new codebook is calculated according to the defined CC (eq.~\ref{eq:centroids}).
	\item Return to step \ref{lbg:partition_calculation}.
\end{enumerate}

\subsubsection{Discussion on VQ applied to our classification problem}
\label{subsubsec:3.2.3}
Key works in the field of VQ have pointed out, both from a theoretical and an experimental point of view, that VQ approaches converge towards partitions whose cells contribute almost equally to the total distortion \cite{Gersho86,Chinrungrueng95}. Under this hypothesis, one can easily state that the nuclear physics classification problem cannot be approached through standard VQ algorithms \cite{Benkirane95,Alderighi01}. In particular, according to the features observed in Fig.~\ref{fig:fig01}, a physically meaningful partition of the of the ($E$,$\Delta E$) plane is characterized by hyperbolic-like cells with strongly different distortions. As an example, the cluster associated to ($Z=1$,$A=1$), indicated with $^{1}$H in Fig.~\ref{fig:fig01}, contains a number of patterns equal to several orders of magnitude those of ($Z=5$, $A=10$), $^{10}$B. Consequently, the distortion introduced by $^{1}$H is significantly larger than that produced by $^{10}$B. Even if a normalization to the number of patterns is introduced, distortions would still be unbalanced among cells as a result of the intrinsically different dispersion of patterns around their mean value observed in different clusters; this aspect is more quantitatively discussed in Section~\ref{sec:sec2}. Another significant limitation to the application of standard VQ algorithms is represented by the need to know the physically meaningful number of classes $N_C$ a priori. The latter, as stated in Section~\ref{sec:sec2}, is usually different for each individual case.

In order to develop a suitable unsupervised learning approach for the classification of nuclear physics data based on VQ, we have modified the original LBG method of Ref.~\cite{Linde80} via the introduction of physical constraints deduced by the formal treatment of the interaction of radiation with the matter \cite{Livingston37}. In our modified LBG, a codebook corresponds to a particular choice of $N_{par}$ physical parameters $P_i$ that allow the calculation of a family of curves (${\cal C}$), one for each physical class ($Z$,$A$), being $N_C$ fixed. Accordingly, the related distance function $d$ is defined in $\Re^2 \times {\cal C}$. The computation of the Voronoi partition, calculated as in eq.~\ref{eq:voronoi}, is used for the NNC. The CC corresponds instead to the determination of the best set of functional parameters $P_i$ for a given partition. The latter is operated by means of a gradient descent technique \cite{Bishop06} applied to the mathematical expression of the total distortion in the parameters space.

Resulting VQ algorithm is used as a hill-climbing operator devoted to the local search for a maximum of the fitness function (and therefore a minimum of the quantization error of the codebook) in the proximity of a good individual determined by an EC approach. To this extent, the initial codebook $Y_0$, composed by the number of suitable physics classes $N_C$, their ($Z$,$A$) values and an initial choice of parameters $P_i$, is uniquely determined, fully automatically, through a global search procedure via evolutionary criteria, and the VQ algorithm plays the role to speed-up the search for the maximum.

The resulting approach is called \emph{Constrained Evolutionary Clustering} (C-EC) and is described in detail in Section~\ref{sec:sec4}. Sections~\ref{sec:sec5} and \ref{sec:sec6} are instead dedicated, respectively, to discuss the capabilities of the algorithm in classifying experimental data and to a detailed comparison with previously published algorithms for the classification of data in nuclear physics experiments at low and intermediate energies.

\section{The Constrained Evolutionary Clustering algorithm}
\label{sec:sec4}
\subsection{Overview of the algorithm}
\label{sec:sec4.overview}
C-EC is an algorithm for automatic data classification in nuclear physics experiments based on EC and VQ. It is conceived for the classification problem typical of experiments that involve the detection of charged particles produced in nucleus-nucleus collisions at low and intermediate energies through longitudinally stacked detectors. In previously published automatic approaches \cite{Benkirane95,Alderighi01,Alderighi01b,Morach08} the link between extracted clusters and meaningful physics classes ($Z$,$A$) was a non-trivial task, often requiring non-negligible human supervision and/or the use of a priori physics information, leading the scientific community to more often rely on human-supervised classification methods \cite{LeNeindre02,Gruyer17}. To overcome these limitations, C-EC combines unsupervised learning techniques with physically meaningful constraints. The resulting solution is a codebook, whose codewords are directly linked to physical classes.
\begin{figure}[t!]
	\centering
	\includegraphics[scale=0.3]{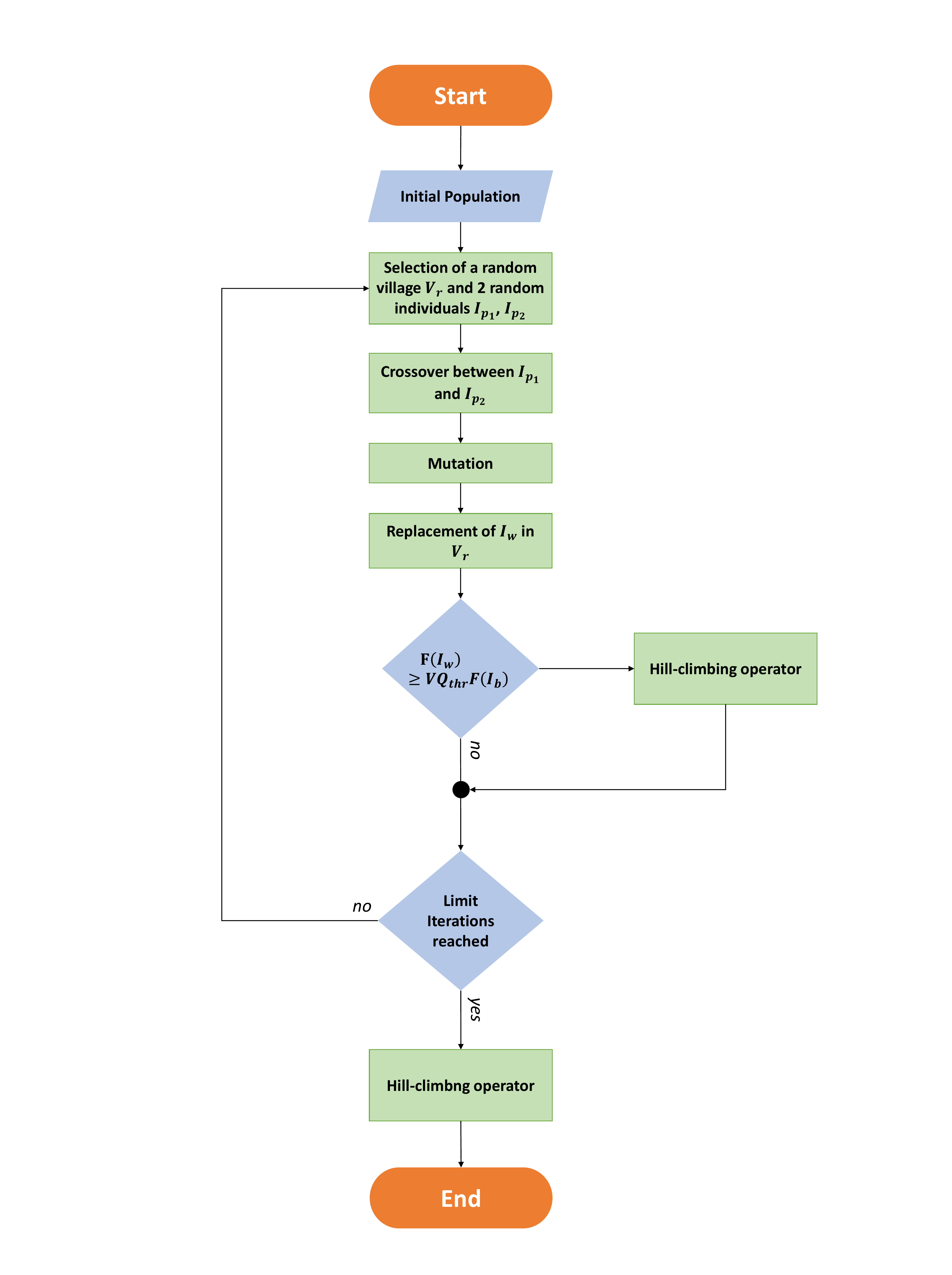}
	\caption{Description of the EC algorithm for initial codebook determination. When the genetic iterations end, i.e. when a pre-defined number of iterations is reached, the VQ block is invoked. The VQ block is also invoked during the genetic iterations, when a particularly promising individual is produced.}
	\label{fig:fig02}
\end{figure}
\begin{figure}[t!]
	\centering
	\includegraphics[scale=0.3]{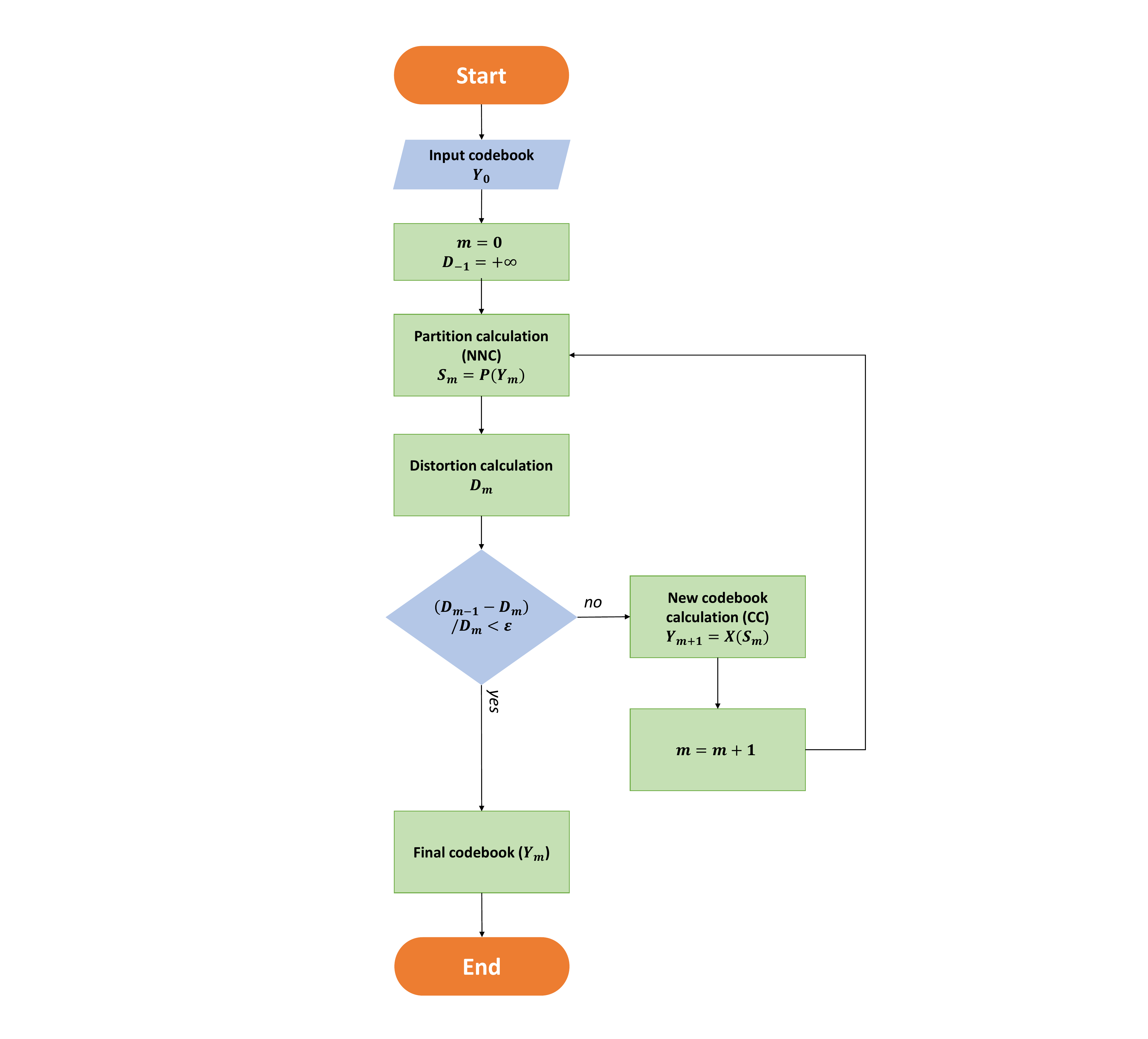}
	\caption{Description of the VQ algorithm devoted to codebook optimization. This block is usually invoked by the EC algorithm, that provides the initial codebook $Y_0$.}
	\label{fig:fig03}
\end{figure}

The algorithm is based on two-level blocks: the upper block is an EC algorithm that is devoted to derive, fully automatically, a physically meaningful codebook for the clusterization problem through the improvement of suitably defined individuals. To enhance the convergence speed, a hill-climbing operator, which deals with the optimization of the codebook via a VQ algorithm, is introduced. The latter consists in the lower block of the algorithm and can be intended as the fast search for a local maximum of the fitness in the proximity of the individual determined by the EC block. Figures~\ref{fig:fig02} and \ref{fig:fig03} report flow charts describing, respectively, the steps executed by the EC and VQ blocks. The VQ algorithm is invoked exclusively by the EC block as a local hill-climbing operator.

\subsection{Upper level: the EC algorithm}
\label{sec:sec4.ec}
\subsubsection{Evolutionary mechanism}
\label{sec:sec4.ec.mechanism}
The mechanism outlined by the flow chart of Fig.~\ref{fig:fig02} can be described as follows:
\begin{enumerate}
	\item \label{ev:individuals_extraction} A village $V_r$ is considered, selected according to a random uniform distribution. Two random individuals, $I_{p_1}$ and $I_{p_2}$, are selected within the village $V_r$ (including overlap regions with neighbor villages). Within the rest of the village $V_r-\{I_{p_1},I_{p_2}\}$, the worse individual, i.e. the individual $I_{w}$ having the lowest fitness value, is identified.
	\item The crossover between $I_{p_1}$ and $I_{p_2}$ is executed, followed by a mutation as described in Section~\ref{sec:sec4.ec.crossover}.
	\item The offspring is introduced in the population replacing $I_w$.
	\item If the fitness of the newly introduced individual is greater than a certain prefixed fraction $VQ_{thr}$ of the fitness of the best individual ($I_b$) in the population, hill-climbing operator is invoked for the optimization of the new individual. The latter (described in Section.~\ref{sec:sec4.vq}) is a particularly time consuming task and therefore $VQ_{thr}$ is usually chosen to be greater than $1$.
	\item If the number of iterations executed is lower than a certain prefixed value, the algorithm returns to step~\ref{ev:individuals_extraction}.
	\item The best individual in the population is optimized by invoking the hill-climbing operator.
\end{enumerate}

\begin{figure}[t!]
	\centering
	\includegraphics[scale=0.6]{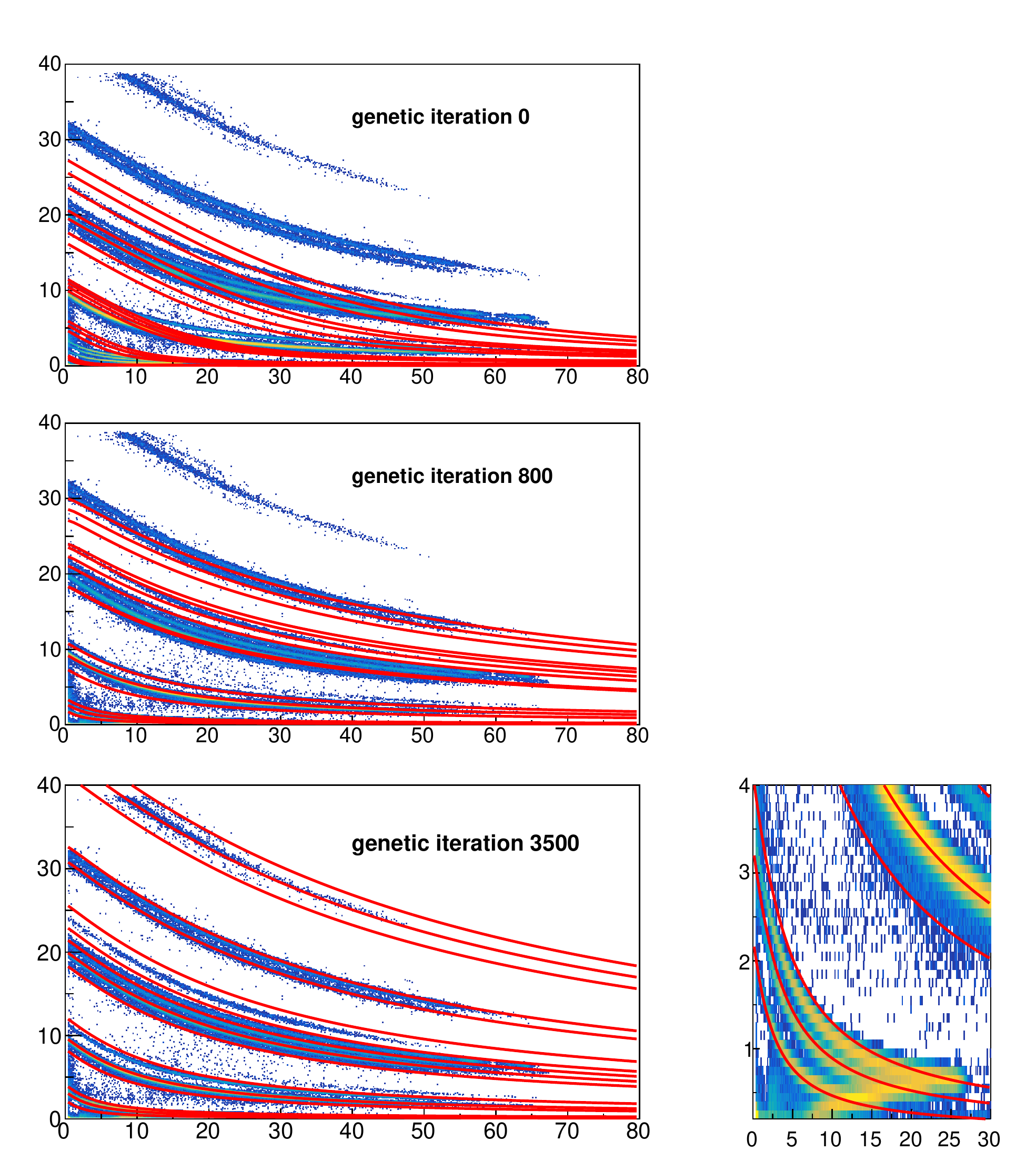}
	\caption{Experimental data (color scale represents the counts in each bin, blue corresponds to less counts) and a visual representation of the best individual in the population after $0$ genetic iterations (top panel), $800$ genetic iterations (middle panel) and $3500$ genetic iterations (bottom panels). The resulting codebook after $3500$ iterations is quite satisfactory to the classification process. The bottom-right panel is a zoom of the low $Z$ clusters after $3500$ iterations.}
	\label{fig:fig08}
\end{figure}
Figure~\ref{fig:fig08} shows an example of individual improvement by adopting a population of $N_v=13$, $N_i=5$, with an overlap of $1$ individual between villages. $3500$ genetic iterations are considered. The best individual in the population after $0$, top panel, $800$, middle panel, and $3500$ iterations, bottom panels (right panel is a zoom of the low-$Z$ region), is shown. Data, represented by points (the color scale represents the counts), are obtained by using a pair of longitudinally stacked silicon detectors (see Section~\ref{sec:sec5} for additional experimental details). The individual is represented by red lines, each corresponding to a given codeword. It is obvious that the results after $0$ iterations, i.e. the best individual in the initial population generated randomly, produces a highly unsatisfactory classification of data. After about $800$ iterations, the result is visually satisfactory for low $\Delta E$ clusters. The solution obtained after $3500$ genetic iterations is very close to a good maximum of the fitness function. By visually inspecting bottom panels in figure, on can clearly see that all clusters are correctly identified, including a group of $3$ poorly populated and high-dispersion clusters located in the upper part of the bi-dimensional distribution. A number of $3500$ iterations is found to be largely sufficient to obtain a fully satisfactory codebook for all cases explored in this paper. It is reasonable that the number of iterations to perform could be slightly adjusted by the experimenter based on the performance of the algorithm in the classification problem explored.
\subsubsection{Genetic encoding of an individual}
\label{sec:sec4.ec.encoding}
The EC algorithm schematically described by the flow chart in Fig.~\ref{fig:fig02} is focused on the global search for an optimal solution of the clusterization problem. This is done through the improvement of individuals via the implementation of the selection criteria described in Sect.~\ref{sec:sec4.ec.mechanism}. Because individuals represent solutions of a clusterization problem, they are suitably encoded to represent codebooks of the data to classify. In a similar way, one can state that the EC block operates a codebook improvement. However, differently from the local search performed by the VQ block (Sect.~\ref{sec:sec4.vq}), which is done in the proximity of the input codebook determined by the EC block, here any possible codebook is explored and the search is in this sense \emph{global}.

As previously stated, one of the major novelties of our approach consists in the use of physical constraints for the derivation of a physically meaningful codebook. We define the adopted codebook in the following way. Let us consider a functional of the form $\Delta E = f_P(E,Z,A)$, where ($E$,$\Delta E$) pairs $\in \Re^2$ represent the coordinates of patterns in the original bi-dimensional distribution to classify, a ($Z$, $A$) pair indicates a given ion and $P$ is a set of $N_{par}$ numerical parameters that constrain the functional. For a given $\bar{P}=\{\bar{P}_0,\ \dots,\ \bar{P}_{N_{par-1}}\}$ set, the vector ($f_{\bar{P}}(E,Z,A)$, $E$) represents the location of a ($Z$,$A$) isotope in the ($E$,$\Delta E$) plane, corresponding to the abscissa $E$ and given the parameter vector $\bar{P}$. A different choice of parameters $P_i$ will result in a different location. If $Z=\bar{Z}$ and $A=\bar{A}$ are fixed, i.e. if a certain ion is considered, the locus of ($f_{\bar{P}}(E,\bar{Z},\bar{A})$, $E$) points can be treated as the codeword associated to the cluster ($\bar{Z}$, $\bar{A}$), being $\bar{P}$ the functional parameters. Without full mathematical rigor, let us indicate with ${\cal C}$ the class of all possible ($f_{{P}}(E,{Z},{A})$, $E$) curves, corresponding to any ($Z$,$A$) isotope and any possible choice of parameter set $\{P_i\in\Re\}$. A codebook can be then defined by considering the $N_C$ elements of ${\cal C}$ corresponding to a certain set of parameters $\bar{P}$ and a certain choice of ($Z$,$A$) pairs $\{(\bar{Z}_1,\bar{A}_1),\ \dots,\ (\bar{Z}_{N_C-1},\bar{A}_{N_C-1})\}$. In such a way, different codewords within a codebook differ exclusively for ($Z$,$A$) values, i.e. each codeword is related to a different ion being the $P$ set fixed. The clusterization problem is therefore equivalent to the search for the optimal set of parameters $P$ and of ($Z$,$A$) pairs.

In this framework, the functional $f_{{P}}(E,Z,A)$ is any possible parametric functional suitable for the description of ($E$,$\Delta E$) signals produced by charged particles in longitudinally stacked detectors. Several valuable models are available in the literature, being derived from the well-know formalism described in \cite{Livingston37}, see for example \cite{TassanGot02,LeNeindre02}. To produce the results described in this paper, we adopted the analytical model introduced in Ref~\cite{TassanGot02}:
\begin{equation}
\begin{split}
\Delta E = & f_P(E,Z,A) = \\  
= & \left[\left(P_0E\right)^{P_1+P_2+1}+\left(P_3Z^{P_4}A^{P_5}\right)^{P_1+P_2+1}+P_6Z^2A^{P_1}\left(P_0E\right)^{P_2}\right]^{\left(1/(P_1+P_2+1)\right)}
\end{split}
\label{eq:tassangot}
\end{equation}
Where the dependence on $Z$ and $A$ is explicit. For the functional of equation \ref{eq:tassangot}, one has $N_{par}=7$. 

\begin{figure}[t]
	\centering
	\includegraphics[scale=0.8]{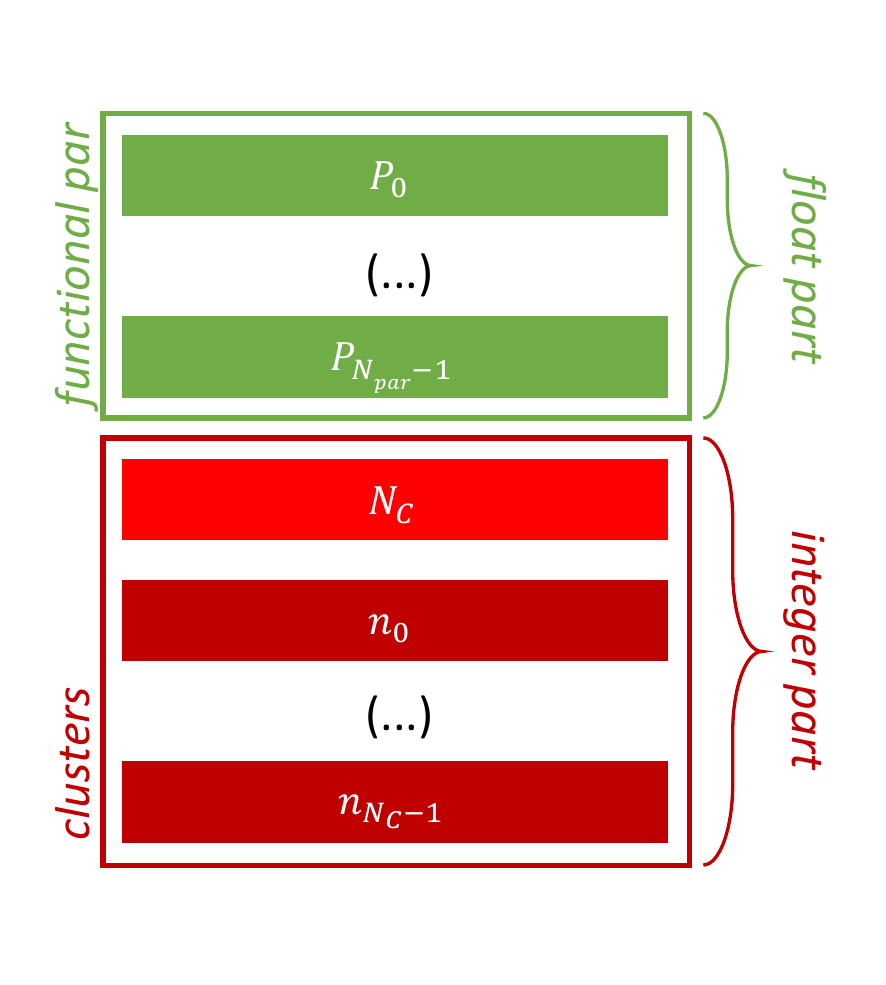}
	\caption{Genetic encoding of an individual according to the EC algorithm developed in this work. A hybrid float/integer encoding is chosen in order to suitably implement a codebook to the clusterization problem.}
	\label{fig:fig04}
\end{figure}
To fulfill the requirements of the physically meaningful codebook described above, we adopted a hybrid float/integer genetic encoding. Figure~\ref{fig:fig04} schematically represents the encoding of an individual. The first $N_{par}$ floating points are used to identify a given parameter set for the functional given by eq.~\ref{eq:tassangot}. Possible ($Z$,$A$) pairs identifying physically meaningful isotopes are considered according to the compilation published on Ref.~\cite{Audi17} and organized into a database for increasing $Z^2A$ values, as suggested by the simplified formula of eq.~\ref{eq:eq1}. In this way, larger indexes are associated to ($Z$,$A$) pairs whose data lie on a higher $\Delta E$ region of the ($\Delta E$,$E$) plane and only one numerical index is needed to identify a given isotope. For the sake of clarity, we have introduced the notation $n_i$ to indicate the index corresponding to the $i$-th cluster, being $i=0,\ \dots,\ N_C-1$. Together with the $N_{par}$ functional parameters, $N_C$ and the set $\{n_i;\ \ i=0,\ \dots,\ N_C-1\}$ are required to complete the encoding of an individual. In this manner, an individual identifies a unique meaningful codebook.
\subsubsection{Population and villages}
\label{sec:sec4.ec.population}
\begin{figure}[t]
	\centering
	\includegraphics[scale=0.6]{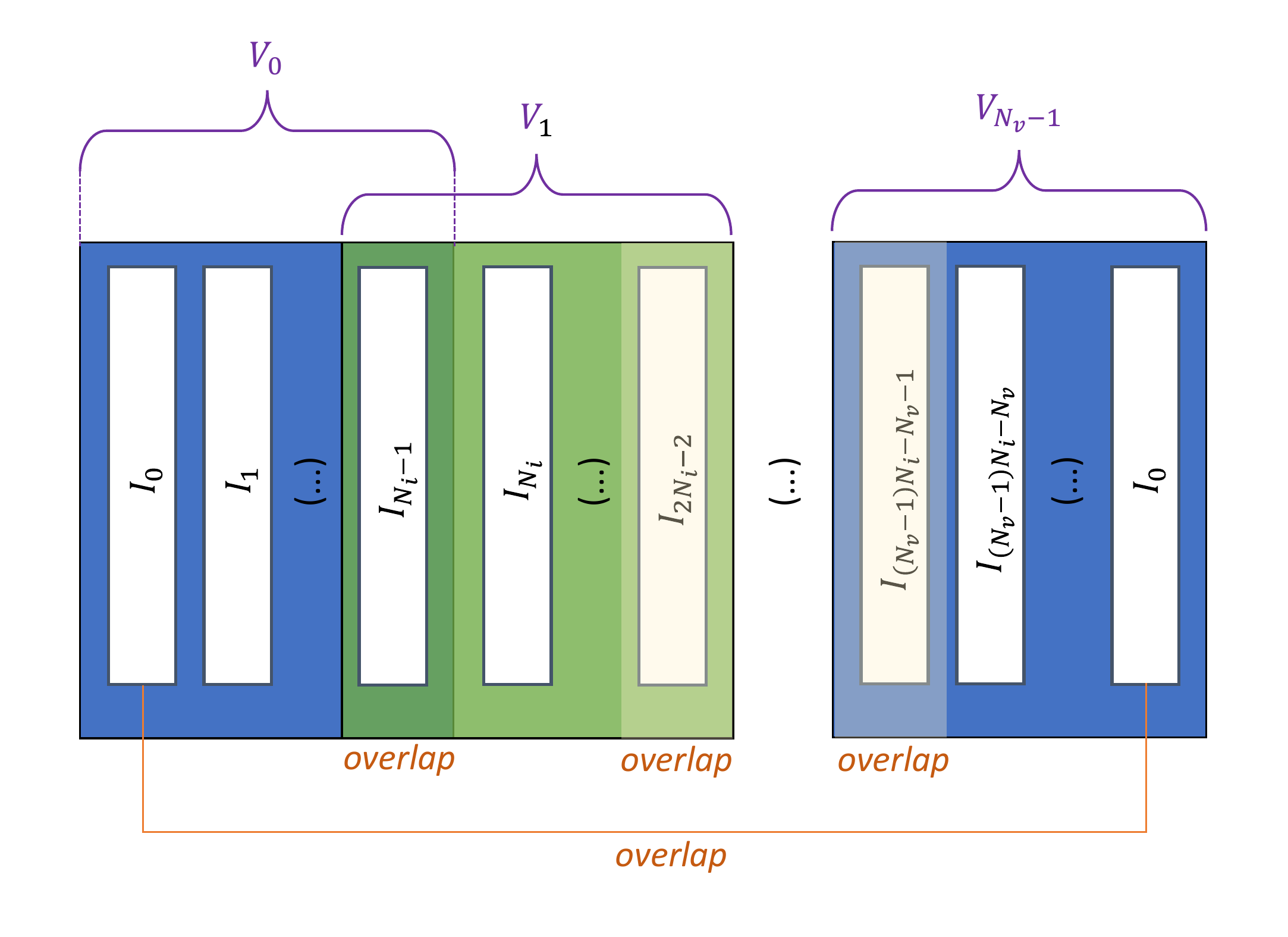}
	\caption{The distribution of individuals in the population.}
	\label{fig:fig05}
\end{figure}

The initial population is generated randomly. Resulting individuals ($I_i$) are distributed according to the scheme described in Fig.~\ref{fig:fig05}. In order to limit the probability of premature convergence of the algorithm and to suitably implement the migration, as discussed in Section~\ref{sec:sec3.1}, the population is subdivided into $N_v$ villages, each containing an equal number $N_i$ of individuals. The first and last individual of each village belong simultaneously to two contiguous villages. In this way, as shown in Fig.~\ref{fig:fig05}, first and last villages in the population are connected. The presence of such overlap regions effectively implements the migration of individuals through different villages. It is important to note that, to avoid premature convergence of the entire population, overlap involves exclusively pairs of neighbor villages. 

The generation process of each individual involves four different steps. (1) A set of functional parameters is produced, being each parameter generated according to a uniform random distribution. For simplicity, parameters are chosen to vary within physically meaningful ranges. The latter can be easily determined by considering the physical meaning of each parameter, as discussed in Ref.~\cite{TassanGot02}. (2) $N_C$ is generated with a uniform integer distribution, with minimum value $1$. (3) $N_C$ numerical indexes $n_i$ are picked randomly without duplicates. (4) The fitness of the newly generated individual is calculated.
\subsubsection{The fitness function}
\label{sec:sec4.ec.fitness}
The selection mechanism operated by the EC algorithm is based on the fitness function, that we indicate with $f_{fit}$, being the latter the objective to maximize. Consequently, a suitable choice of $f_{fit}$ is crucial to the success of the algorithm. We define three terms that contribute in the fitness function: $f_e$, related to total the distortion error associated to the codebook identified by the individual, $f_n$, related to the number of codewords $N_C$, and $f_{avg}$, related to the average distortion per unit of pattern associated to each $Z$-group of clusters.

The total distortion is obtained through the following equation:
\begin{equation}
  e = \sum_{i=0}^{N_P-1}{d({\bf x}_i,q({\bf x}_i))}
  \label{eq:total_distortion}
\end{equation}
where $d:\Re^2\times{\cal C} \rightarrow \Re^+_0$ is the modified distance function:
\begin{equation}
  d({\bf x}_i,q({\bf x}_i))=|{\Delta E}_i-f_P(E_i,Z,A)|
  \label{eq:distance}
\end{equation}
being ${\bf x}_i=(E_i,{\Delta E}_i)$ the $i$-th pattern and $q({\bf x}_i)=\{(E,f_P(E,Z,A))\}$ the codeword associated to the $i$-th pattern within the considered codebook. The latter is easily determined by scanning through the codebook till the codeword with the minimum $|{\Delta E}_i-f_P(E_i,Z,A)|$ value is found.

By using the formulation of the total distortion introduced by eq.~\ref{eq:total_distortion}, the $f_e$ term can be defined as follows:
\begin{equation}
  f_e = \frac{e_{max}-e}{e_{max}}
  \label{eq:f_e}
\end{equation}
where $e_{max}$ is the maximum expected error a priori. By default, $e_{max}$ is calculated as the mean absolute error of the pattern ordinates: $e_{max}=\frac{1}{N_P}\sum_{i=0}^{N_P}{|{\Delta E}_i-{\Delta E}_{avg}|}$.

From the formulation of $f_e$, it is obvious that introducing a larger number of ($Z$,$A$) clusters in the codebook represented by the individual usually contributes to decrease the distortion error $e$ and therefore to increase $f_e$. For this reason, if only $f_e$ contributes to $f_{fit}$, the algorithm will naturally converge towards individuals with numerous clusters; $f_n$ is required to contrast this phenomenon. We define $f_n$ in the following way:
\begin{equation}
  f_n = \frac{n_{max}-N_C}{n_{max}}
  \label{eq:f_n}
\end{equation}
where $n_{max}$ is the maximum allowed number of clusters.

$f_{avg}$ is introduced in the fitness function to account for the following facts pointed out in Section~\ref{sec:sec2}: (i) the number of patterns within a cluster differs significantly between clusters, (ii) clusters associated to higher $Z$-values have larger dispersion and the dispersion is roughly similar between different sub-clusters of a given $Z$-value. For $f_{avg}$ one has:
\begin{equation}
  f_{avg}=\frac{1}{N_Z}\sum_{Z^{\prime}}\sum_{A^{\prime}}\frac{1}{N_P^{(Z^\prime,A^\prime)}}{D}_{(Z^{\prime},A^\prime)}
  \label{eq:f_avg}
\end{equation}
where we have indicated with $N_{Z}$ the number of different $Z$-values present in the codebook, the sum on $Z^\prime$ is extended to all available $Z$-values, the sum on $A^\prime$ is extended to all ($Z^\prime$,$A^\prime$) sub-clusters corresponding to the same $Z^\prime$ value and ${D}_{(Z^{\prime},A^\prime)}$ is the total distortion (eq.~\ref{eq:total_distortion}) introduced by the cluster ($Z^\prime$,$A^\prime$), being $N_P^{(Z^\prime,A^\prime)}$ its total number of patterns. Equation~\ref{eq:f_avg} has the meaning of average distortion per pattern of $Z$-groups of clusters. The distortion per pattern is used to account for the requirement (i), outlined above, while averaging on all $Z$-values has a physical meaning because of (ii).

The terms of equations~\ref{eq:f_e}, \ref{eq:f_n} and \ref{eq:f_avg} are combined together to form the fitness function $f_{fit}$:
\begin{equation}
  f_{fit}=100\frac{f_eu(f_e)+\alpha_nf_nu(f_n)+\alpha_{avg}f_{avg}}{1+\alpha_n+\alpha_{avg}}
  \label{eq:f_fit}
\end{equation}
Where we introduced the Heaviside step function $u(x)$ to smooth the effects due to solutions with $e>e_{max}$ or $N_C>n_{max}$. $\alpha_n$ and $\alpha_{avg}$ are used to tune the relative importance of $f_n$ and $f_{avg}$, respectively, with respect to $f_e$. Typically, $\alpha_n\in[0.01,0.02]$ and $\alpha_{avg}\in[0.1,0.5]$ are found suitable for the classification process. 
\subsubsection{Crossover and mutation}
\label{sec:sec4.ec.crossover}
\begin{figure}[t]
	\centering
	\includegraphics[scale=0.34]{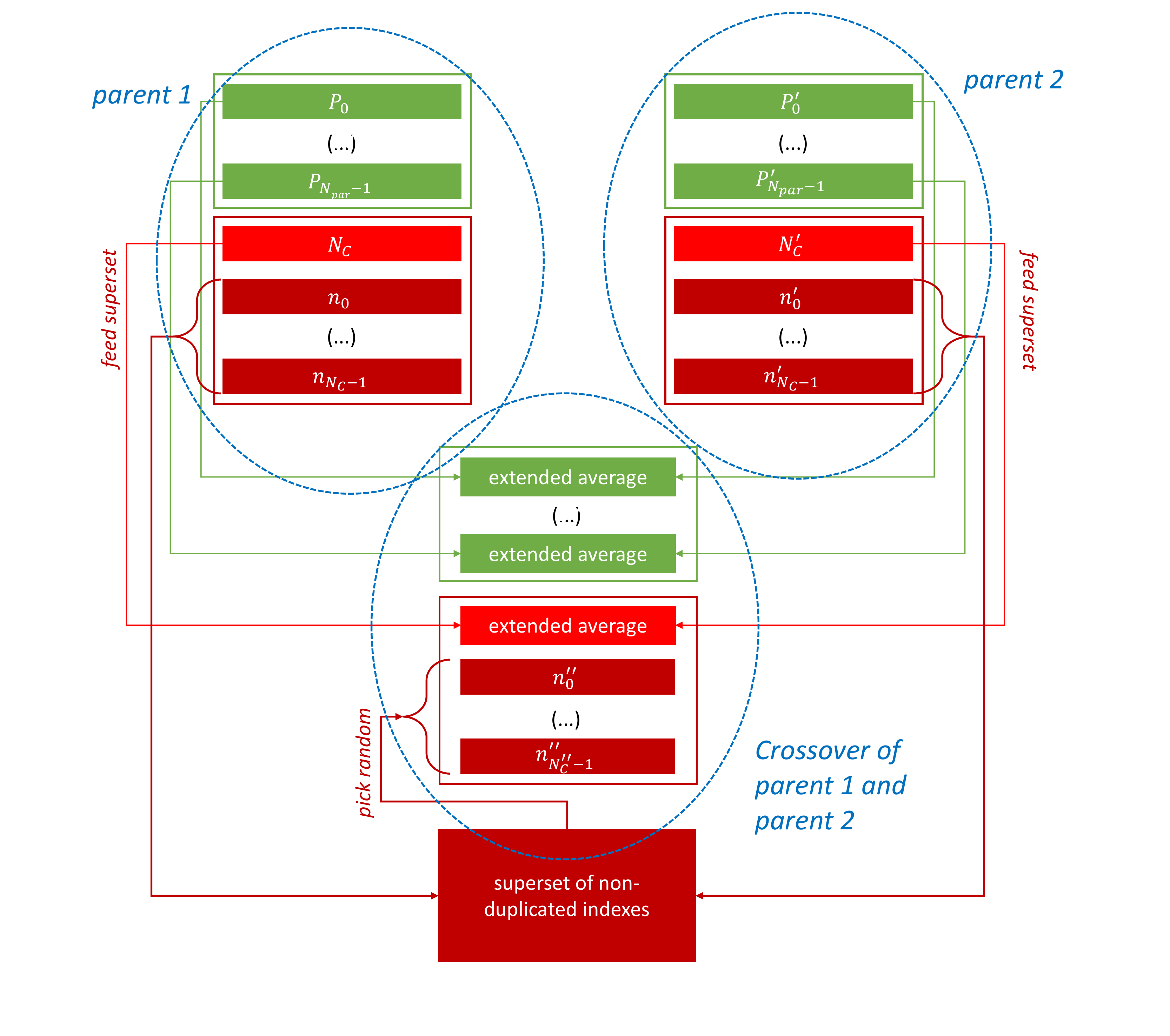}
	\caption{A schematic description of the crossover between two individuals.}
	\label{fig:fig06}
\end{figure}
Crossover is operated through the algorithm schematically described in Fig.~\ref{fig:fig06} and according to the following procedure. Let us consider two individuals, $I$ and $I^\prime$, $I$ having encoding $\{P_0,\ \dots,\ P_{N_{par}-1}\}$, $N_C$, $\{n_0,\ \dots,\ n_{N_C-1}\}$ and $I^\prime$ with encoding $\{P^\prime_0,\ \dots,\ P^\prime_{N_{par}-1}\}$, $N^\prime_C$, $\{n^\prime_0,\ \dots,\ n^\prime_{N^\prime_C-1}\}$. In addition, let us indicate with $I^{\prime\prime}$, $\{P^{\prime\prime}_0,\ \dots,\ P^{\prime\prime}_{N_{par}-1}\}$, $N^{\prime\prime}_C$, $\{n^{\prime\prime}_0,\ \dots,\ n^{\prime\prime}_{N^{\prime\prime}_C-1}\}$, the result of the crossover of $I$ and $I^\prime$. Tho distinct processes are performed for the functional parameters and the clusters. Each functional parameter $P^{\prime\prime}_i$ of the offspring $I^{\prime\prime}$ is obtained via a so-called \emph{extended average} of the analogous parameters from the encoding of $I$ and $I^\prime$:
\begin{equation}
  P^{\prime\prime}_i = \alpha P_i + (1-\alpha)P^{\prime}_i
  \label{eq:extended_par}
\end{equation}
where $\alpha$ is a random number (a newly generated for each $i=0,\ \dots,\ N_{par}-1$) uniformly generated in the range $\alpha\in[-\varepsilon_{par},1+\varepsilon_{par}]$, being $\varepsilon_{par}\approx0.15$ a certain constant.

A similar procedure is performed for the number of clusters: $N^{\prime\prime}_C = \alpha N_C + (1-\alpha)N^{\prime}_C$, $\alpha$ being extracted in the range $\alpha\in[-\varepsilon_{C},1+\varepsilon_{C}]$. Once $N^{\prime\prime}$ is determined, a \emph{superset} of indexes $n$ is produced containing all the non-duplicated indexes that better approximate the original $I$ and $I^\prime$ codewords in the new parameter set $\{P^{\prime\prime}_i\}$. $N^{\prime\prime}_C$ indexes are then picked randomly from the superset of indexes. If the size of the superset is lower than $N^{\prime\prime}_C$, the remaining indexes to complement the set $\{n^{\prime\prime}_i\}$ are extracted randomly from the database constructed using the data of Ref.~\cite{Audi17}. In a similar way, one obtains an individual with the following characteristics: (i) functional parameters $P^{\prime\prime}_i$ have intermediate values between those of the parents, but some probability to obtain external values exists, (ii) the number of codewords is close to those of the parents but some probability to be larger or smaller than that of the parents exists, (iii) the spatial disposition of resulting codewords is in good geometrical matching with codewords of the parents $I$ and $I^\prime$, even if, with some small probability, clusters in regions of the ($E$,$\Delta E$) plane not covered by $I$ and $I^\prime$ codewords can be produced.
\subsubsection{Mutation and genetic diversity monitoring}
Mutation is implemented both for the float and integer part of the individual encoding. For the float part, the, each of the functional parameters is altered with a small probability $p_{mut}^{par}$, usually chosen to range within $0.001$ and $0.04$. For the integer part, resulting $N^{\prime\prime}_C$ is altered of one unity ($\pm1$) with a small probability $p_{mut}^{C}\in[0.001,0.07]$.

An interesting mechanism of $p_{mut}^{par}$ and $p_{mut}^{C}$ variation is introduced to ensure the stability of the genetic diversity, thus helping to contrast the premature convergence of the algorithm. At the beginning, $p_{mut}^{par}$ and $p_{mut}^{C}$ are the minimum possible. This choice is to allow a fast elimination of particularly bad individuals. To monitor the genetic diversity, the ranges of parameter variation are then divided in $100$ cells for each of the parameters. Each cell corresponds to a given choice of a given parameter, within a small interval. A cell is considered full if at least one individual has a parameter contained in the corresponding interval. At each iteration of the EC block, the total number of non-empty cells is calculated. This value is expressed as percentage of the total number of cells and is used to evaluate the genetic diversity of the population. If the genetic diversity drops below a prefixed threshold, $p_{mut}^{par}$ is increased. On the contrary, $p_{mut}^{par}$ is suitably decreased if the genetic diversity surpasses a certain threshold. 
\subsubsection{Smart insertion and elimination}
After the process summarized by Fig.~\ref{fig:fig06} is done, an algorithm of smart elimination and insertion of codewords is executed. This deals with removing unnecessary codewords and inserting more useful codewords. Unnecessary codewords are identified by calculating the number of patterns in the corresponding cluster. If the latter is lower than a certain minimum size, the codeword is removed from the codebook. New codewords are then inserted in order to restore their original number. Insertion process is executed according to the following steps: (i) a loop through all the codewords is done starting from a randomly selected codeword, (ii) if a cluster whose distortion per unit of pattern is above the average in the codebook is found, the indexes related to its lower and upper neighbor clusters are taken into consideration. (iii) If one of those is not present in the codebook, it is then introduced. If after this procedure the number of clusters is lower than the original number of clusters before smart elimination, indexes are picked randomly from the database of ions (without duplicates) till the required number is reached.

\begin{figure}[t]
	\centering
	\includegraphics[scale=0.6]{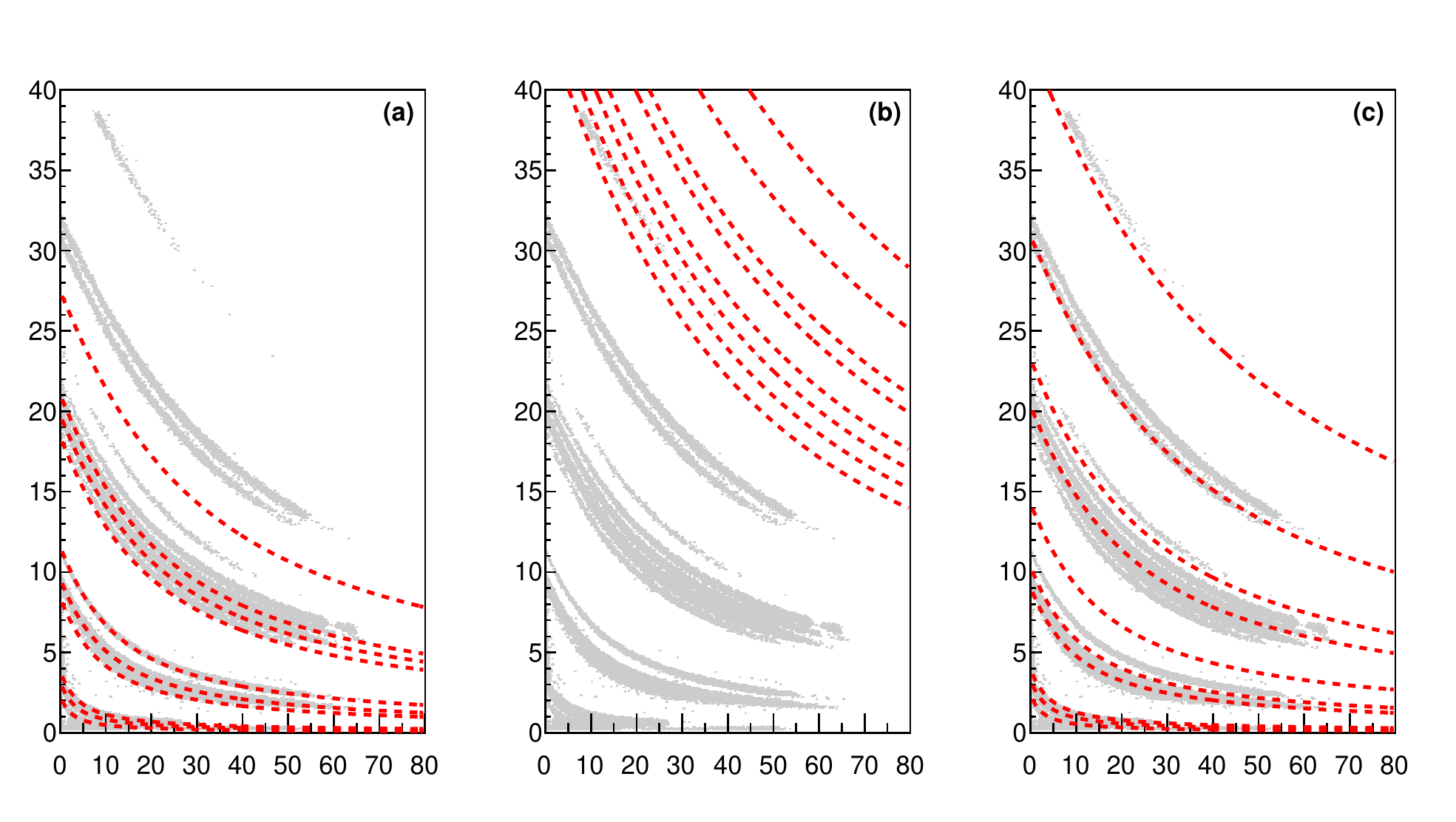}
	\caption{An example of crossover (c) between the individuals (a) and (b). Experimental data are represented by gray dots. The codebook encoded by the individuals is graphically represented by means of red dashed lines. Each panel shows one individual. Resulting individual (c) contains clusters whose geometrical dispositions is in matching both with some (a) and (b) codewords.}
	\label{fig:fig07}
\end{figure}
Figure \ref{fig:fig07} shows a graphical example of crossover between two individuals. In the figure, experimental data are represented by gray dots, while the codebook encoded by individuals is represented by dashed red lines. Data are obtained by using two longitudinally stacked silicon detectors, as described in Section~\ref{sec:sec5}. Individuals are indicated with labels (a), (b) and (c). A single panel is used to represent each individual: (a) Left panel, (b) center panel, (c) right panel. (c) represents the result of the crossover between (a) and (b). In this example, (a) covers only the lower region of the ($E$,$\Delta E$) plane, while, on the contrary, (b) contains uniquely clusters geometrically located in the upper part of the plane, resulting in a highly unsatisfactory classification for low $Z$ clusters. In addition, a number of clearly unnecessary clusters is present, especially in the codebook identified by the individual (b). The resulting crossover (c), obtained with the prescriptions discussed above and by combining the parents (a) and (b), contains codewords whose geometrical disposition covers suitably the entire range of data. In addition, the number of clusters that are unnecessary to the classification process is reduced, as a result of the smart elimination and insertion process. 

\subsection{Lower level: local hill-climbing operator via a modified LBG algorithm}
\label{sec:sec4.vq}
By carefully inspecting the bottom panels of Fig.~\ref{fig:fig08}, one can observe a number of codewords not in fully satisfactory agreement with observed clusters. This is due to the fact that a large number of iterations is required for an EC algorithm to fully converge to an absolute maximum of the fitness function, and the performances are consequently poorer when approaching extremely high fitness individuals. To speed-up the optimization of the codebook, the VQ block is used as hill-climbing operator for the EC block.

Codebook optimization is operated through the algorithm schematically described in Fig.~\ref{fig:fig03}. After the initial codebook $Y_0$ is calculated (output of the EC block), the following optimization steps are iterated:
\begin{enumerate}
	\item \label{vq:nnc_calculation} Iteration $m$-th begins. The Voronoi partition $S_m$  of the input data is computed (NNC, Sect.~\ref{sec:sec4.vq.nnc}) according to the $Y_m$ partition calculated in the $(m-1)$-th iteration.
	\item The total distortion $D_m$ associated to the $S_m$ partition is calculated (Section~\ref{sec:sec4.vq.mqe}).
	\item If $(D_{m-1}-D_m)/D_{m}<\varepsilon$, being $\varepsilon$ a given prefixed value, then $Y_m$ is used as final codebook and the optimization process ends.
	\item A new codebook $Y_{m+1}$ is calculated (CC, Section~\ref{sec:sec4.vq.cc}) starting from the partition determined in step~\ref{vq:nnc_calculation}.
	\item The algorithm returns to step~\ref{vq:nnc_calculation}.
\end{enumerate}
\subsubsection{NNC}
\label{sec:sec4.vq.nnc}
The NNC consists in the calculation of the Voronoi partition ${\cal P}(Y)=\{\bar{S}_1,\ \dots,\ \bar{S}_{N_C}\}$ according to the definition of eq.~\ref{eq:voronoi}. Adopted distance function is that of eq.~\ref{eq:distance}.
\subsubsection{MQE}
\label{sec:sec4.vq.mqe}
MQE is defined by eq.~\ref{eq:dist_cod}, being the total distortion within a cell $D_i$ calculated according to eq.~\ref{eq:tot_cella} and by using the modified distance function defined by eq.~\ref{eq:distance}.
\subsubsection{CC}
\label{sec:sec4.vq.cc}
Let us assume that a codebook $\bar{Y}$ is defined by the parameter set $\bar{P}=\{\bar{P}_0,\ \dots,\ \bar{P}_{N_{par}-1}\}$, a number of codewords $\bar{N}_C$ and the set of physics classes $\{(\bar{Z}_0,\bar{A}_0),\ \dots,\ (\bar{Z}_{\bar{N}_C-1},\bar{A}_{\bar{N}_C-1})\}$. If one considers an input partition $\bar{S}=\{\bar{S}_0,\ \dots,\ \bar{S}_{\bar{N}_C-1}\}$, where $\bar{S}_i$ is the cell associated to the physics class $(\bar{Z}_i,\bar{A}_i)$, CC consists in a suitable variation of each individual parameter $P_i$ ($i=0,\ \dots,\ N_{par}-1$) in order to minimize the MQE associated to $\bar{Y}$. It is important to note that, according to the scheme introduced in this paper, a variation of any parameter $\bar{P}_{i}$ will affect the absolute position of all codewords in the codebook, in a physically meaningful way. To this end, such MQE minimization corresponds to the search for the minimum of the total distortion function in the parameters space.

The function of the total distortion error in the parameters space can be defined as follows:
\begin{equation}
  E(P_0,\ \dots,\ P_{N_{par}-1})=\sum_{i=0}^{N_{C}-1}\sum_{j=0}^{N_P^i-1}|\Delta E_j^i-f_P(\Delta E_j^i,Z_i,A_i)|
  \label{eq:distortion_par_space}
\end{equation}
where the sum on $i$ is extended to all the cells in the input partition, the sum on $j$ is intended on all the $N_P^i$ patterns in the $i$-th cell, and ($Z_i$, $A_i$) is the physics class associated to the $i$-th cell. The dependence on the parameters $\{P_i\}$ is implicit in the definition of $f_P$ as shown, for example, for the model defined by eq.~\ref{eq:tassangot}. CC consists in the minimization of the function $E(P_0,\ \dots,\ P_{N_{par}-1})$. To this end, a gradient descent technique is used \cite{Bishop06}.

The mathematical expression of the gradient of $E(P_0,\ \dots,\ P_{N_{par}-1})$ in the parameters space can be easily derived from eq.~\ref{eq:distortion_par_space}:
\begin{equation}
\nabla E(P_0,\ \dots,\ P_{N_{par}-1})= 
  \begin{bmatrix}
   \frac{\partial E}{\partial P_0} \\
   \dots \\
   \frac{\partial E}{\partial P_{N_{par}-1}}
  \end{bmatrix}
\label{eq:gradient_par_space}
\end{equation}
and involves the partial derivatives of $f_P$ with respect to $P_k$, $\frac{\partial f_P}{\partial P_k}$. The latter can be computed analytically for a functional $f$ like the one of eq.~\ref{eq:tassangot}, but numerical approximations can also be used if more complex functionals are adopted. The vector defined by eq.~\ref{eq:gradient_par_space} represents the direction, in the parameters space, of maximum increment for $E$. The direction identified by $-\nabla E$ can be therefore used to vary the parameter vector $P$ towards the maximum decrement of $E$.

$E(P_0,\ \dots,\ P_{N_{par}-1})$ minimization can be thus executed through the following steps:
\begin{enumerate}
	\item A vector $\eta=\{\eta_0,\ \dots,\ \eta_{N_{par}-1}\}$ is defined, being $\eta_i$ suitably small numbers. The latter depend usually on the amplitude of the error function gradient. For the model of eq.~\ref{eq:tassangot}, values $\eta_i\approx10^{-7}$ are found appropriate. However, he use of non-optimal initial $\eta_i$ values will not affect the accuracy of the results but exclusively the number of iterations required for the minimization of $E$.
	\item \label{cc:init_error} $E_{prev}=E(P_0,\ \dots,\ P_{N_{par}-1})$ is calculated and conserved.
	\item \label{cc:p_new} A new set of parameters is obtained starting from the initial parameter vector $P$ and according to the opposite of the direction of the gradient, $P_i^{new}=P_i-\eta_i(\nabla E(P_0,\ \dots,\ P_{N_{par}-1}))_i$.
	\item \label{cc:error_new} $E_{new}=E(P^{new}_0,\ \dots,\ P^{new}_{N_{par}-1})$ is calculated and conserved.
	\item if $E_{new} > E_{prev}$, then all $\eta_i$ are multiplied by a suitably small factor, usually $\approx0.1$, and the algorithm returns to step \ref{cc:p_new}.
	\item \label{cc:eta_new} The gradient in the new parameter set $\nabla E(P^{new}_0,\ \dots,\ P^{new}_{N_{par}-1})$ is calculated. If $\nabla E(P^{new}_0,\ \dots,\ P^{new}_{N_{par}-1})_i\nabla E(P_0,\ \dots,\ P_{N_{par}-1})_i>0$ then $\eta_i$ is multiplied by a factor greater than $1$, usually $\approx1.1$, otherwise it is multiplied by a suitably small factor, usually $\approx0.1$.
	\item \label{cc:update_parameters} The parameters are updated to the new values $P_{i}=P_{i}^{new}$.
	\item Steps \ref{cc:init_error} to \ref{cc:update_parameters} are iterated until the condition $(E_{prev}-E_{new})/E_{new}<\varepsilon_{CC}$ is verified, being $\varepsilon_{CC}$ a prefixed value.
	\item The minimum for $E(P_0,\ \dots,\ P_{N_{par}-1})$ is found in correspondence of the parameter set $\{P_0,\ \dots,\ P_{N_{par}-1}\}$ and the CC is terminated.
\end{enumerate}
\section{Test of the algorithm with experimental data}
\label{sec:sec5}
To probe the capabilities of the C-EC algorithm in the classification of nuclear physics data, we considered experimental data recorded by two longitudinally stacked layers of silicon. The experiment was performed at the TRIUMF laboratory of Vancouver (Canada). A $^{9}$Li accelerated beam was delivered on a LiF target at an energy of $65$ MeV. $^{9}$Li+$^{6}$Li, $^{9}$Li+$^{7}$Li and $^{9}$Li+$^{19}$F collisions were investigated. They resulted in the production of several ions especially in the range $1\leq Z\leq 5$. Detection apparatus consisted of $6$ pairs of semiconducting detectors (silicon) each having $16$ individual detection units. Consequently, the number of independent bi-dimensional distributions to classify is $96$. The experiment was aimed at the investigation of the production of resonances in light neutron-rich nuclei. The technique was based on the exploration of the invariant mass of ions emitted in resonance disintegration events. In similar studies, that are widely used in experiments focused at the discovery of new resonances, high-precision data classification is critical. As an example, the experiment explored the existence of new highly-excited states of $^{12}$Be and their decay in channels involving the emission of clusters such as $^{6}$He+$^{6}$He, $^{8}$He+$^{4}$He, $t$+$^{9}$Li or $p$+$^{11}$Li. Such information, and especially distinguishing among the various emission channels, is relevant to fully understand the production of clustered states in nuclei \cite{Charity07,Freer01} and the possible formation of nuclear molecules \cite{Kanada03}. Despite the relatively low complexity of the apparatus, our experiment represents a good benchmark for automatic classification methods because of the requirement of a highly-precise identification of a variety of ions, including both $Z$ and $A$ values.

\begin{figure}[t]
	\centering
	\includegraphics[scale=0.55]{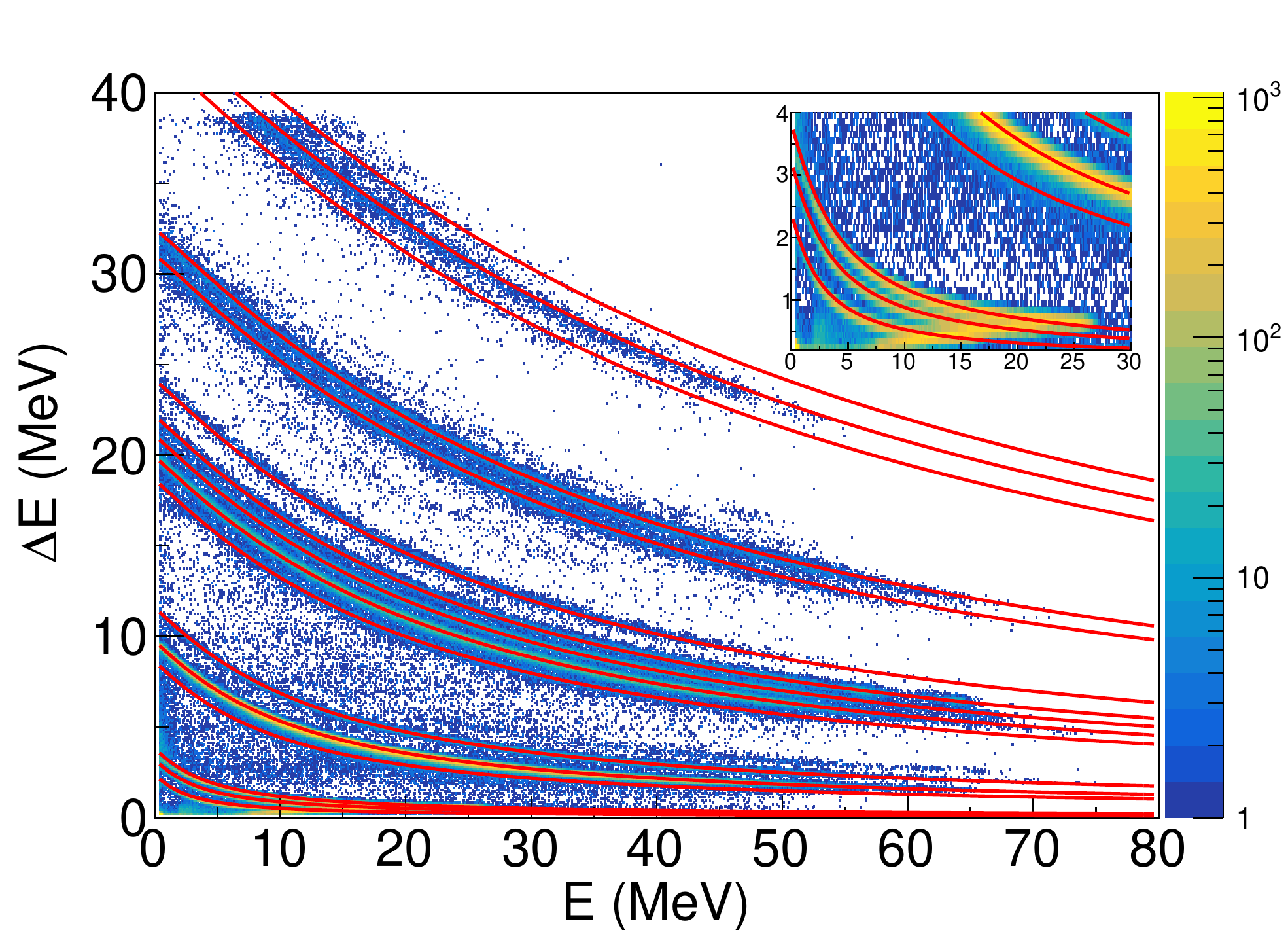}
	\caption{Result of the classification of data collected in the TRIUMF experiment for a typical detector. Data are represented by dots, whose color reflects the number of counts in a given bin. Deduced codebook used for the classification (red lines represent the position of each codeword) is produced after $3500$ genetic iterations and the VQ algorithm optimization process. The insert shows a zoom of the low $Z$ region.}
	\label{fig:fig09}
\end{figure}
Figure~\ref{fig:fig09} shows the result of the C-EC processing of a typical detector case after $3500$ genetic iterations and local search operated by the VQ block. If compared to the bottom panels of Figure~\ref{fig:fig08}, where the best codebook after $3500$ genetic iteration is shown, one can semi-quantitatively observe that the various codewords significantly better approximate observed clusters with respect to the pure EC individual improvement. This a result of the codebook optimization procedure. The result is extremely satisfactory for all cluster. The insert in figure shows a zoom of the low $Z$ region. Even if a deviation in the high $E$ part of clusters is present, reflecting ions that have sufficient kinetic energy to punch-through the second detection layer, and resulting in an inversion of the cluster towards low $\Delta E$ and $E$ values, low $Z$ codewords are in fully satisfactory agreement with clusters corresponding to $^{1}$H, $^{2}$H and $^{3}$H isotopes. A similar deviation is present in many experiments in this energy domain and corresponding data populate a region undesired to the classification process. Furthermore, none of the identified codewords is associated to regions populated by background counts present in the spectrum. This is a particularly relevant result as it shows that the algorithm is almost insensitive to noise.

\begin{figure}[t!]
	\centering
	\includegraphics[scale=0.55]{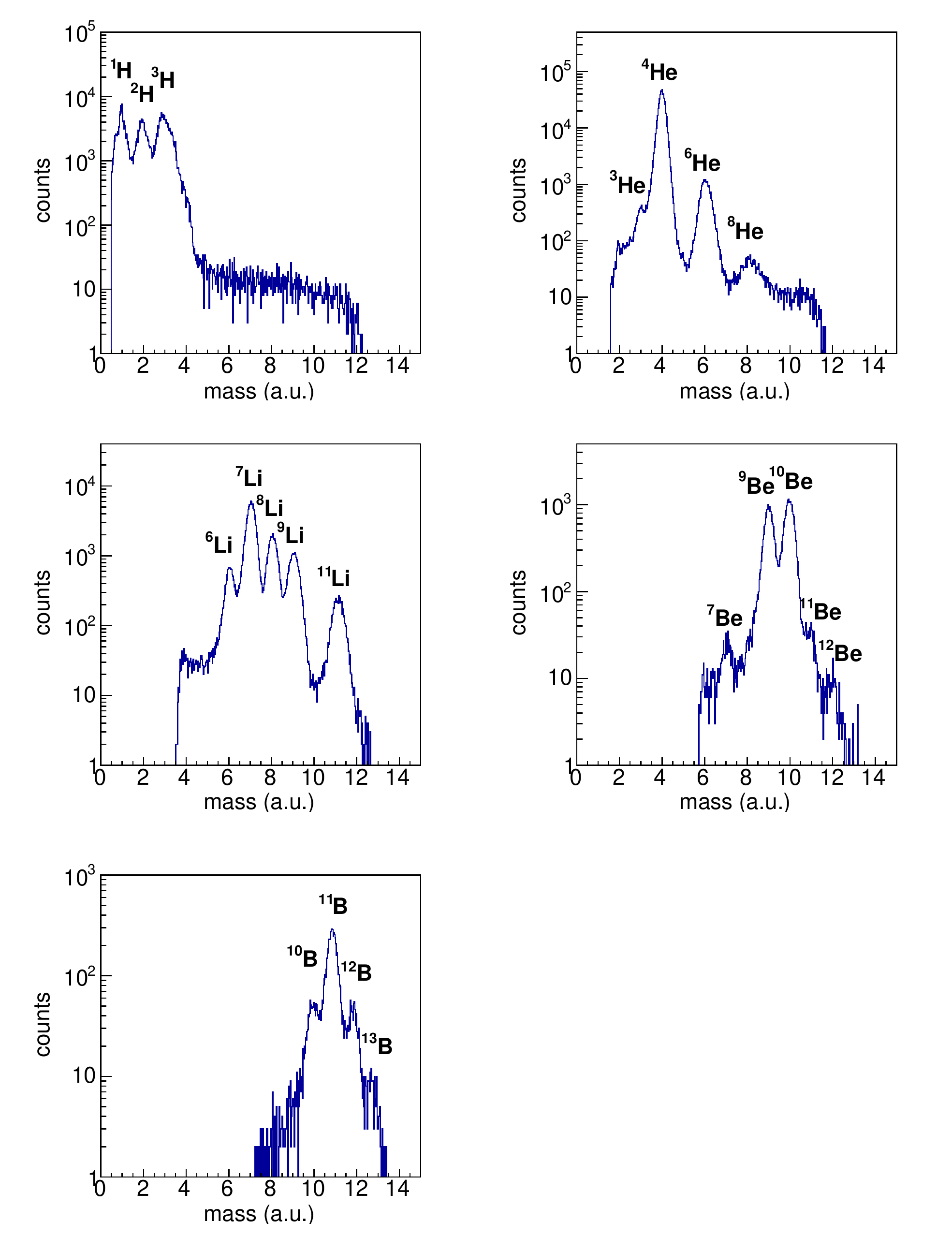}
	\caption{Quantitative analysis of the quality of the classification obtained from the codebook shown in Fig.~\ref{fig:fig09}. We adopted a similar procedure than the one proposed in Ref.~\cite{LeNeindre02}. Data are subdivided into $5$ different panel, each corresponding to an identified $Z$-value. Peaks in the panels correspond to each identified $A$-value associated to a given $Z$, they are identified by labels.}
	\label{fig:fig10}
\end{figure}
The plot of Figure~\ref{fig:fig09} allows to a visual semi-quantitative inspection of the produced codebook but does now allow a completely quantitative analysis of the quality of the classification. Classification is quantitatively proven through a procedure similar to the one proposed in Ref.~\cite{LeNeindre02}. Data shown on Figure~\ref{fig:fig09} is initially subdivided into groups corresponding to their identified $Z$-value. This is done by considering their proximity to groups of codewords associated to each $Z$-value. If a ($\Delta \bar{E}$, $\bar{E}$) point is closer to the group of lines associated to a certain $\bar{Z}$, then ($\Delta \bar{E}$, $\bar{E}$) is classified as $Z=\bar{Z}$. By considering all data classified to a given $\bar{Z}$-group, then a point is associated to the closest possible $A$ for that particular $\bar{Z}$-value, $A=\bar{A}$. To quantify the quality of the classification, the normalized distance from the $\bar{A}$ codeword to the point is calculated. The distance is normalized in such a way that the neighbor codeword ($A=\bar{A}+1$) has a distance of $1$ from the $\bar{A}$-codeword. In this way, a point is classified as ($\bar{Z}$,$\bar{A}$) if it has a normalized distance lower than $0.5$ from the ($\bar{Z}$,$\bar{A}$) codeword. In Figure~\ref{fig:fig10} we report normalized distance distributions for data classified as $Z=1,2,3,4,5$, respectively from left to right and from top to bottom. Several peaks are present and associated to different identified isotopes from H to B. Peaks associated to $^{1}$H, $^{2}$H and $^{3}$H are well-visible, even if they are broader as a result of the unavoidable effect of punch-through discussed above. Very interestingly, unphysical classifications are not present. As an example, no peak is visible at $A=5$ for $Z=2$, as the corresponding $^{5}$He is an unbound nucleus. Similarly, as expected by a meaningful classification, $^{10}$Li and $^{8}$Be peaks are also missing. The latter is characterized by an extremely reduced lifetime (of the order of $10^{-16}$ seconds) and undergoes therefore to decay into lighter fragments before reaching the detectors.

\begin{figure}[t!]
	\centering
	\includegraphics[scale=0.55]{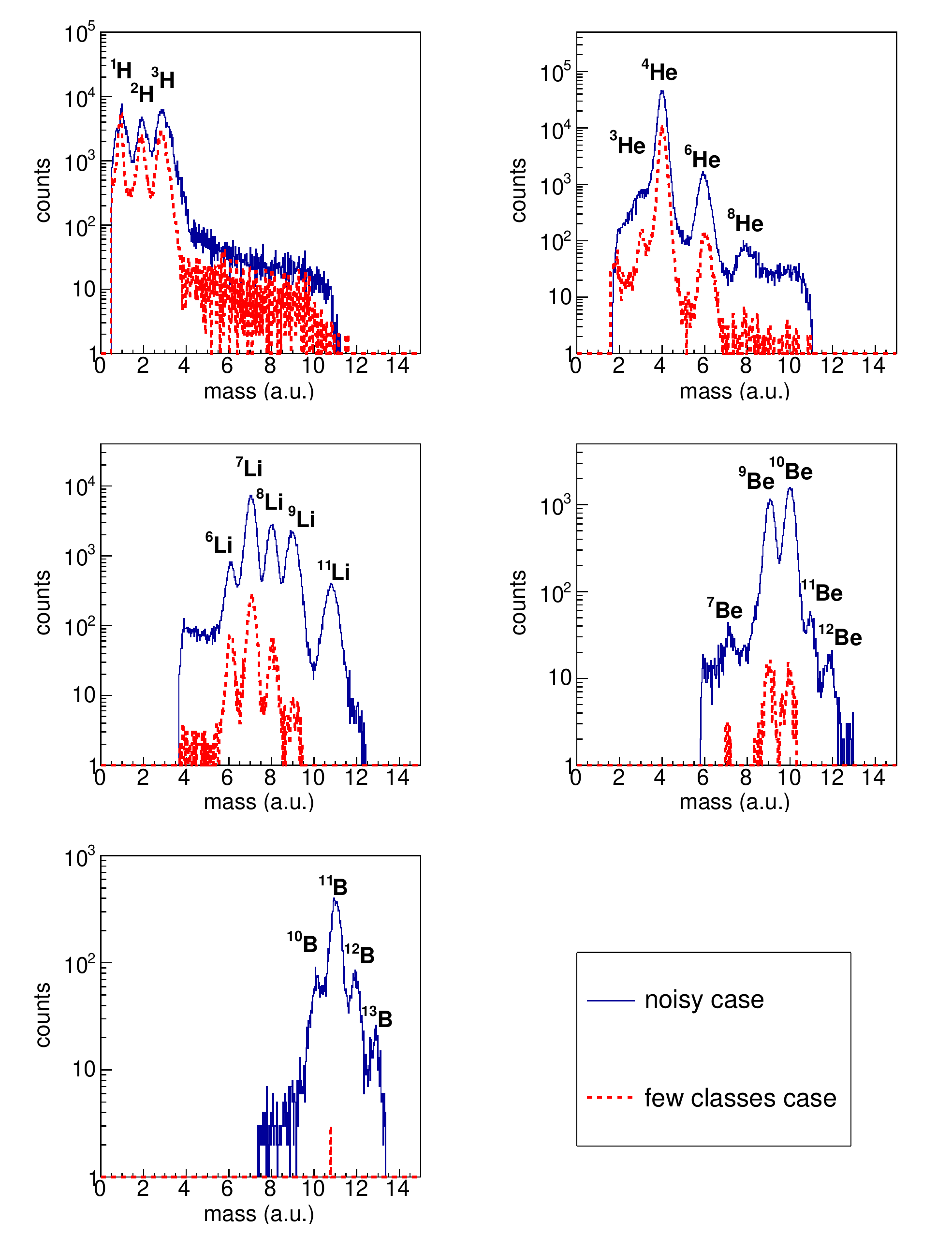}
	\caption{Analysis of the classification results, with the procedure suggested in Ref.~\cite{LeNeindre02}, for a particularly noisy case (solid line) and a detector characterized by the presence of a reduced number of clusters (dashed red line). The first is a detector placed close to the beam line; in this case larger statistics of particles in a broader energy and $Z$ domain are expected. Data for the few clusters case is collected by a detector placed at a larger distance from the beam line; lower rate of ions in a narrower $Z$ domain is expected in this case. Obtained classification is highly satisfactory in both cases.}
	\label{fig:fig13}
\end{figure}
Extremely satisfactory classifications have been obtained, consistently, for all $96$ detectors explored in our benchmark. Figure~\ref{fig:fig13} shows the obtained classifications, according to the method proposed in Ref.~\cite{LeNeindre02}, for a particularly noisy detector (solid lines in panels of Fig.~\ref{fig:fig13}) and for a detector characterized by a reduced number of clusters (dashed red line in Fig.~\ref{fig:fig13}). Differences in the detectors arise essentially from their geometrical disposition: the noisy detector was placed very close to the beam line and is therefore subject to a larger rate of impinging ions having broader energy distributions, causing the presence of undesired noise, while the other is placed at a larger distance from the beam line, being therefore characterized by a lower rate of incident ions in a narrower $Z$ domain. As clearly observed, meaningful physics classes are correctly identified in both cases. In the top left panel ($Z=1$), the three peaks associated to $^{1}$H, $^{2}$H and $^{3}$H are present, but they are broader for the noisy detector, reflecting a larger amount of data in the region corresponding to ion punch-through. The peak corresponding to $^{3}$He in the top right panel ($Z=2$) results partially merged with the neighbor $^{4}$He peak (populated with the highest statistics) in the case of the noisy detector, while the $^{3}$He-$^{4}$He separation is excellent for the few clusters case. $^{8}$He peak is extremely weakly populated in the few clusters case, reflecting the low statistics of production of the neutron-rich ion $^{8}$He for the emission direction covered by the detector. A more populated $^{8}$He peak is instead observed for the noisy case, where the statistics of production of a larger number of ions is quite significant. Very interestingly, as expected, $Z$ range is more limited for the detector placed at a larger distance from the beam line, and, in particular, no clusters are identified as $Z=5$. Also in the $Z=4$ classes the two detectors differ, one can observe significant $^{7}$Be, $^{9}$Be, $^{10}$Be, $^{11}$Be and $^{12}$Be peaks in the detector placed close to the beam line, while only $^{9}$Be and $^{10}$Be are significantly identified for the one distant from the beam line.

The analysis discussed in this section comprises $96$ independent classification problems. The CPU time requested to complete the task resulted to be about $440$ seconds on a commercial Intel i7-9700K (8 cores) processor at a frequency of $3.4$ GHz. This is a satisfactory result, which testifies that the algorithm is suitable for analysis of data produced by large apparatus (where the number of individual classification problems is usually between several hundreds and a few thousands).
\section{Comparison with other approaches}
\label{sec:sec6}
Let us summarize the capabilities of the C-EC algorithm: (i) no a priori information is required, number of clusters to classify and relevant ($Z$,$A$) values are obtained through an individual improvement algorithm based on evolutionary criteria, (ii) after suitable codebook optimization, the solution of the clusterization problem is readily usable for data classification with explicit link to physically meaningful classes, (iii) the algorithm is robust to noise.

This section is dedicated to a detailed comparison between the C-EC algorithm and previous approaches published in the literature \cite{Benkirane95,Alderighi01,Alderighi01b,Morach08,Wirth13,IaconoManno00,LeNeindre02,Gruyer17}.

The work published in Ref.~\cite{Benkirane95} is probably the first example of classification of nuclear physics data recorded by stacks of detectors via unsupervised learning approaches. It is based on the visual analysis of bi-dimensional assembly of data via sophisticated image segmentation techniques. This approach has been used exclusively in cases where only $Z$-classification is meaningful. Major limitation of the approach consists in the need for a priori information such as the slope of clusters and their inter-distances. Even if the authors show how to systematically calculate these quantities based on physical considerations within $1\%$ accuracy, the method is effective only for a reduced class of detectors for which a reliable energy calibration is possible without assumptions on data classification.

Ref.~\cite{Alderighi01} is concerned with the use of pre-attentive neural systems. Performances are particularly good for groups of clusters characterized by similar inter-distance and dispersion, e.g. the high $Z$-region, where the resolution is not sufficient to distinguish sub-clusters associated to different $A$-values. The effectiveness of the approach in the low $Z$-region, where one can expect to observe a separation of clusters due to the $A$-value, is reduced as a result of the variety of dispersion and inter-distance between clusters. The link to physically meaningful classes relies on the definition of $Z$-value for the last identified cluster.

\begin{landscape}
	\begin{table}[]
		\centering
		\begin{tabular}{ccccccc}
			\hline
			Work                     	  & approach         &  sub-clusters&  $Z$-range     & low-energy & no a priori         & link to   \\
			~                             &                  &  class.      &                & class.     & info.               & ($Z$,$A$) \\
			\hline
			Ref.~\cite{Benkirane95}       &  image           &  no          & full           & yes        & no                 & from a priori info          \\ 
			~                             &  segmentation    &              &                &            &                     &           \\ \hline
			Ref.~\cite{Alderighi01}       &  pre-attentive   &  no          & only high $Z$  & yes        & yes                 &  no\\ 
			~                             &  neural system   &              &                &            &                     & \\ \hline
			Ref.~\cite{Alderighi01b}      &  spatial density &  no          & full           & no         & yes                 &  no\\
			~                             &  data processing &              &                &            &                     & \\ \hline
			Ref.~\cite{Morach08}          &  data slicing    &  yes         & full           & yes        & yes                 &  no\\
			~                             &                  &              &                &            &                     & \\ \hline
			Ref.~\cite{Wirth13}           &  fuzzy           &  -           & -              & yes        & yes                 &  no\\ 
			~                             &  c-means         &              &                &            &                     & \\ \hline
			Ref.~\cite{IaconoManno00}     &  artificial      &  yes         & full           & yes        & no                  &  from a priori info\\ 
			~                             &  neural networks &              &                &            &                     & \\ \hline
			Ref.~\cite{LeNeindre02}       &  modeling of     &  yes         & full           & yes        & no                  &  from a priori info\\ 
			~                             &  patterns        &              &                &            &                     & \\ \hline
			Ref.~\cite{Gruyer17}          &  data slicing    &  yes         & full           & yes        & no                  &  from a priori info\\ 
			~                             &                  &              &                &            &                     & \\ \hline
			This work              		  &  EC and VQ       & yes          & full           & yes        & yes                 & yes\\ 
			~                             &   				 &    		    &  				 & 			  & 					& \\ \hline
		\end{tabular}
		\caption{Summary of the comparisons with other approaches for nuclear physics data classification.}
		\label{tab:comparisons}
	\end{table}
\end{landscape}
A spatially density data processing is reported in Ref.~\cite{Alderighi01b}. The procedure is based on the identification of points along clusters and their subsequent linearization. Performances are strongly reduced in the low energy part of data, i.e. low $E$ by looking at Fig.~\ref{fig:fig01}, because of the rapid change in slope of clusters. In addition, sub-clusters associated to $A$-values are not identified. The procedure is particularly powerful for online data analysis cases, where a reduced body of information is typically sufficient, for example to monitor the stability of large detectors.

More recently, Ref.~\cite{Morach08} has proposed a classification method based on a simple data slicing procedure. The approach is capable to identify both $Z$-clusters and $A$-clusters but the association of extracted clusters to physically meaningful ($Z$,$A$) classes is left to the operator.

Fuzzy clustering methods derived from the c-means were adopted in Ref.~\cite{Wirth13}. However, the approach is restricted exclusively to cases comprising few classes.

Artificial neural networks were adopted in Ref.~\cite{IaconoManno00} to approximate the spatial disposition of physics clusters in bi-dimensional distributions of data. The underlying mechanism is that of a supervised learning approach, where a number of patterns are manually extracted by the operator for each ($Z$,$A$) class to then feed the training process of the neural network. Even if resulting clusters are directly linked to ($Z$,$A$) pairs, the whole procedure heavily relies on human-supervision.

In a similar way with respect to Ref.~\cite{IaconoManno00}, the authors of Ref.~\cite{LeNeindre02} adopted a supervised learning procedure where the classification capabilities rely on a number of manually extracted patterns. Differently from Ref.~\cite{IaconoManno00}, a formal model of data with explicit ($Z$,$A$) dependence is used, thus allowing the extraction of patterns only for a reduced number of classes. Results are extended to all possible ($Z$,$A$) via model extrapolation. The procedure critically relies on the quality of the information extracted by the operator.

Finally, a new approach based on data slicing was very recently proposed in Ref.~\cite{Gruyer17}. The procedure combines peak finding algorithms with information that needs to be manually extracted by an operator. The mechanism is similar to that proposed in Ref.~\cite{LeNeindre02} with the major difference that the patterns needed to constrain the model are extracted automatically from an initial information given by the operator. The advantage of this method with respect to \cite{LeNeindre02} consists therefore in the strongly reduced information required by the algorithm, thus significantly minimizing human supervision. Even if an explicit link to ($Z$,$A$) classes is provided, the result of the classification is affected by the initial information provided by the operator and an a priori visual inspection of data distributions is always required.

Comparisons are summarized in table~\ref{tab:comparisons}. As one can easily observe, our approach is the only one where an explicit link between identified clusters and ($Z$,$A$) values is done in a fully automatic way. The vast majority of other approaches is exclusively concerned with the extraction of clusters and human supervision is consequently required to associate them to physically meaningful classes. A link to $Z$-classes is done by the algorithm of Ref.~\cite{Benkirane95}, but the procedure requires the use of a priori information. It is important to stress that an automatic association is usually a non-trivial task, especially when $A$-classification is required. In a similar case, as observed in the example shown in Fig.~\ref{fig:fig13}, for a given $Z$, not all $A$-sub-clusters are populated and differences can also be observed detector-by-detector. Differently from our approach, the methods of Refs.~\cite{Benkirane95,Alderighi01,Alderighi01b} where not concerned with $A$-cluster extraction. This is probably the case because the algorithms of Refs.~\cite{Benkirane95,Alderighi01,Alderighi01b} were developed for the previous generation of detectors, where $A$-classification was not a crucial requirement. The only algorithms concerned withy explicit link to $Z$ and $A$ classification are those of Refs.~\cite{IaconoManno00,LeNeindre02,Gruyer17}. In all these cases, the quality of the classification crucially relies on the capability of the operator to manually extract the required information via visual inspection of the distribution of data.

\section{Conclusions and perspectives}
\label{sec:sec7}
The classification of data in nuclear physics experiments is key to extract the required physics information. Modern experiments have been largely focused on obtaining high-quality classification over a number of physically meaningful classes. In charged particles experiments at low and intermediate incident energies, the classification problem is that of identifying charge and mass of ions produced in nucleus-nucleus collision. This task is particularly repetitive and time consuming as it usually requires the supervision of an operator that visually inspects and extracts information from bi-dimensional assembly of data.

In this framework, we developed a fully automatic algorithm for data classification in nuclear physics experiments by combining Evolutionary Computing and Vector Quantization. In our approach, a two-level search for the individual with the maximum fitness value is operated. The EC block represents the upper level and deals with a global search extended to all possible individuals, applying suitable evolutionary criteria based on the Darwinian scheme. Once an individual close to a maximum of the fitness function is determined, a hill-climbing operator is invoked to perform a fast local search of the maximum. The latter consists in a modified LBG algorithm that performs a codebook optimization procedure, being fixed the number of codewords and their physical meaning. Our procedure is innovative as it combines unsupervised learning approaches with suitable physics constraints. Resulting solutions are in the form of a codebook, where the link between codewords and physics classes is explicit with nearly-zero effort required to the experimenter. Furthermore, no a priori information is used.

The newly developed algorithm is benchmarked against experimental data obtained by pairs of longitudinally stacked silicon detectors. A satisfactory classification is obtained for all explored cases, including cases with a reduced number of clusters or characterized by noise.

With respect to previously published approaches, our method offers the advantage of the explicit link between extracted clusters and physically meaningful classes, thus significantly simplifying the subsequent analysis of data. The procedure does not rely on any a priori information. This makes the resulting procedure fully-automatic and a minimal human-supervision is only required to inspect the result of the classification

In addition, C-EC is particularly suitable to be integrated in the online and offline analysis of nuclear physics experiments at low and intermediate incident energies, thus allowing a significant reduction of the time required to the analysis of data, especially in large acceptance arrays.




\bibliographystyle{elsarticle-num}
\biboptions{sort&compress}
\bibliography{bibliography_20200211}







\end{document}